\documentclass[journal]{IEEEtran}
\pdfoutput=1
\usepackage{amsmath,amsthm,amsfonts,amscd,amssymb,graphics,dsfont}
\usepackage[mathscr]{eucal}
\usepackage{graphics,graphicx,multicol}
\usepackage{epsfig}
\usepackage{epstopdf}
\usepackage{subfigure}
\usepackage{stfloats}
\usepackage{latexsym}
\usepackage{amsfonts}
\usepackage{cite}
\usepackage{algorithm,algorithmic}
\usepackage{color}

\newtheorem{prop}{Proposition}

\newtheorem{rem}{Remark}

\begin{document}
\title{Deep Learning for Distributed Optimization: Applications to Wireless Resource Management}
\author{Hoon Lee,~\IEEEmembership{Member,~IEEE}, Sang Hyun Lee,~\IEEEmembership{Member,~IEEE}, and Tony Q. S. Quek,~\IEEEmembership{Fellow,~IEEE}
\thanks{This work was supported in part by Institute of Information \& Communications Technology Planning \& Evaluation (IITP) grant funded by the Korea government (MSIT) (2016-0-00208, High Accurate Positioning Enabled MIMO Transmission and Network Technologies for Next 5G-V2X (vehicle-to-everything) Services) and in part by the SUTD-ZJU Research Collaboration under Grant SUTD-ZJU/RES/05/2016 and the SUTD AI Program under SGPAIRS1814. This paper was presented in part at IEEE International Conference on Communications 2019,
Shanghai, China, May 2019 \cite{HLee:19b}. {\em (Corresponding
author: Sang Hyun Lee.)}

H. Lee is with the Department of Information and Communications Engineering, Pukyong National University, Busan 48513, South Korea (e-mail: hlee@pknu.ac.kr).

S. H. Lee is with the School of Electrical Engineering, Korea University, Seoul 02841, South Korea (e-mail: sanghyunlee@korea.ac.kr).

T. Q. S. Quek is with the Singapore University of Technology and Design, Singapore 487372, and also with the Department of Electronic Engineering, Kyung Hee University, Yongin 17104, South Korea (e-mail: tonyquek@sutd.edu.sg).}
}\maketitle 
\begin{abstract}
This paper studies a deep learning (DL) framework to solve distributed non-convex constrained optimizations in wireless networks where multiple computing nodes, interconnected via backhaul links, desire to determine an efficient assignment of their states based on local observations. Two different configurations are considered: First, an infinite-capacity backhaul enables nodes to communicate in a lossless way, thereby obtaining the solution by centralized computations. Second, a practical finite-capacity backhaul leads to the deployment of distributed solvers equipped along with quantizers for communication through capacity-limited backhaul.
The distributed nature and the non-convexity of the optimizations render the identification of the solution unwieldy. To handle them, deep neural networks (DNNs) are introduced to approximate an unknown computation for the solution accurately. In consequence, the original problems are transformed to training tasks of the DNNs subject to non-convex constraints where existing DL libraries fail to extend straightforwardly. A constrained training strategy is developed based on the primal-dual method. For distributed implementation, a novel binarization technique at the output layer is developed for quantization at each node. Our proposed distributed DL framework is examined in various network configurations of wireless resource management. Numerical results verify the effectiveness of our proposed approach over existing optimization techniques.
\end{abstract}
\begin{IEEEkeywords}
Deep neural network, distributed deep learning, primal-dual method, wireless resource management.
\end{IEEEkeywords}

\section{Introduction}
A non-convex optimization has been a fundamental challenge in designing wireless networks owing to the distributed computation nature and the limited cooperation capability among wireless nodes. For several decades, there have been significant efforts in non-convex programming as well as its distributed implementation \cite{MHong:16,YSun:17,Boyd:10,FRKschi:01,Boyd:04}. Although the convergence and the optimality of these techniques have been rigorously examined, the underlying assumptions on network configurations and special mathematical structure of the objective and constraint functions are mostly ideal. For instance, a successive approximation framework \cite{MHong:16} for a non-convex problem requires a proper convex approximation, which usually lacks a generic design technique. Moreover, distributed optimization techniques, such as the dual decomposition, the alternating direction method of multipliers (ADMM) \cite{Boyd:10}, and the message-passing (MP) algorithm \cite{FRKschi:01}, are suitable only for separable objective and constraint functions, and the exchange of continuous-valued messages.
Furthermore, the assumption of ideal cooperation among the nodes can be challenging in practical wireless networks, where any coordination among wireless nodes are usually limited by bandwidth and power constraints.


\subsection{Related Works}
To overcome the drawbacks of traditional optimization techniques, deep learning (DL) frameworks have been recently investigated in wireless resource management  \cite{HSun:18,WLee:18a,PKerret:18,WLee:18b,WLee:18c,WLee:18d} and end-to-end communication system design \cite{OShea:17,Dorner:18,HLee:18a,HLee:18b,HLee:19a}. In particular, ``learning to optimize'' approaches in \cite{HSun:18,WLee:18a,PKerret:18,WLee:18b,WLee:18c,WLee:18d} have received considerable attention for their potential to replace traditional optimization algorithms with neural network computations. Power control problems are addressed in multi-user interference channels (IFCs) via deep neural networks (DNNs) to maximize the sum rate performance \cite{HSun:18}. A supervised learning technique is used to learn a locally optimal solution produced by the weighted minimum-mean-square-error (WMMSE) algorithm \cite{QShi:11}. Its real-time computational complexity is shown to be much smaller than the original WMMSE, albeit involving intensive computations with numerous samples in training. However, the supervised learning task needs the generation of numerous training labels, i.e., the power control solution of the WMMSE algorithm, which would be a bottleneck for the DNN training step. In addition, the average sum rate performance achieved by the DNN in \cite{HSun:18} is normally lower than the WMMSE algorithm due to the nature of the supervised learning framework.

Unsupervised learning strategies, which do not require a label set in training, have been recently applied for power control solutions in the IFC \cite{WLee:18a,PKerret:18}, cognitive radio (CR) networks \cite{WLee:18b}, and device-to-device communication systems \cite{WLee:18c,WLee:18d}. A recent work studies a generic resource allocation formulation in wireless network and develops a model-free training strategy  \cite{MEisen:18}. DNNs are trained to maximize the network utility directly, e.g., the sum rate performance, rather than to memorize training labels. Unlike the supervised counterpart developed in \cite{HSun:18}, training labels are not necessary. In \cite{WLee:18a}, the sum rate maximization over the IFC is pursued by the use of a convolutional neural network (CNN) which accepts channel state matrices of all users as DNN inputs. The designed CNN improves a locally optimal WMMSE algorithm with much reduced computational complexity. Constrained power control problems in a CR network are addressed using DNN so that unlicensed transmitters satisfy interference temperature (IT) constraint for a licensed receiver \cite{WLee:18b,WLee:18c}. However, training DNNs with the IT constraint is not straightforward since existing DL optimizers are not suitable for handling constrained problems. To resolve this issue, a penalizing strategy is applied to transform the original training task to an unconstrained training one by augmenting a penalty term associated with the IT constraint into the network utility function \cite{WLee:18b,WLee:18c}. Nevertheless, the feasibility of the trained DNN is, in general, not guaranteed since it is sensitive to the choice of penalty parameters. Additional optimization of penalty parameters is required for fine-tuning the contribution of the penalizing function in the DNN cost function. The identification of such hyper parameters incurs a computationally expensive search in typical DL applications. Therefore, the method would not be able to obtain the solution of general constrained optimizations in practical wireless communication systems.

Most studies on the DL approach to wireless network optimization require global network information, such as perfect knowledge of channel state information (CSI) at other nodes, and thus the centralized computation process is essential for the operation of the DNN \cite{HSun:18,WLee:18b,WLee:18c,WLee:18d,MEisen:18}. It is not practical due to the limited cooperation capacity among nodes. A simple distributed implementation is investigated in \cite{WLee:18a} for the IFC. Unknown other cell CSI inputs are replaced with zeros in testing the DNN, which is trained in a centralized manner using the global CSI of all transmitters. Due to its heuristic way of the operation and the lack of the other cell CSI knowledge, this simple distributed DNN technique shows a noticeable performance degradation as compared to existing optimization algorithms \cite{WLee:18a}. An on-off power control problem in the
IFC is addressed with individual transmitters equipped with their own DNN for the distributed computation in \cite{PKerret:18}. To enhance the distributed computational capability, the DNN is constructed to accept the estimated CSI for other cells and to yield the on-off transmission strategy of each transmitter. A dedicated DNN is distributed to an individual transmitter after combined training of multiple DNNs. However, it lacks the optimization of the CSI estimation and exchange process, which is a key feature of the distributed network construction. Furthermore, the constrained optimization cannot be handled by the method developed in \cite{WLee:18a} and \cite{PKerret:18}. Therefore, the distributed DL-based design for wireless networks remains open.

\subsection{Contributions and Organization}
This paper investigates a DL-based approach that solves generic non-convex constrained optimization problems in distributed wireless networks. The developed technique establishes a general framework which can include the works in \cite{HSun:18,WLee:18a,PKerret:18,WLee:18b,WLee:18c,WLee:18d} as special cases. Compared to a supervised learning technique \cite{HSun:18}, the proposed DL framework is based on an unsupervised learning strategy that does not require training labels, i.e., the predetermined optimal solutions. There are multiple nodes which desire to minimize the overall network cost by optimizing their networking strategy subject to several constraints. An individual node observes only local measurement, e.g., the local CSI, and quantizes to forward it to neighbors through capacity-limited backhaul links. The quantized information of other nodes can further improve the distributed computing capability of individual nodes. To achieve the network cost minimization, a distributed solver as well as a quantizer of the local information are necessary. This involves a highly complicated formulation with non-convex binary constraints.

To handle this issue, a DNN-based optimization framework is proposed for two different cases according to the capacity of the backhaul links. An ideal infinite-capacity backhaul is taken into account where lossless data sharing is allowed among the nodes. In this configuration, individual nodes can determine their solutions by themselves using the perfect global information collected from other nodes. Thus, a single central DNN unit is designed to produce the overall network solution by collecting the global information as an aggregated input. However, state-of-the-art DL algorithms are mostly intended for unconstrained problems and lack the feature to include constraints in the DNN training. The Lagrange dual formulation is employed to accommodate constrained DNN training problems where the strong duality \cite{Boyd:04} is verified. To train the DNN efficiently under generic network constraints, this work bridges a gap between the primal-dual method in traditional optimization theory and the stochastic gradient descent (SGD) algorithm in the DL technique \cite{LeCun:15}. In consequence, a constrained training strategy is developed such that it performs iterative updates of the DNN and the dual variables via state-of-the-art SGD algorithms. Unlike unconstrained DL works in \cite{HSun:18,WLee:18a,PKerret:18,WLee:18b,WLee:18c,WLee:18d}, the proposed constrained training algorithm ensures to produce an efficient feasible solution for arbitrarily given constraints.

In a more realistic setup of a finite-capacity backhaul, a distributed deployment of the DL framework is addressed. To be specific, an individual node is equipped with two different DNN units: a quantizer unit and an optimizer unit. The quantizer unit generates a quantized version of the local information so that a capacity-limited backhaul afford the communication cost for the transfer of the data to a neighboring node. Meanwhile, the optimizer unit evaluates a distributed solution of an individual node based on the local information along with the quantized data received from other nodes. This distributed DL approach involves additional binary constraints for the quantization unit, resulting in so-called a \textit{vanishing gradient} issue during the DNN training step. Such an impairment has been dealt with by a stochastic operation and a gradient estimation in image processing \cite{YBengio:13,Raiko:15}. For the extension of these studies, we apply a binarization technique of the quantizer unit to address a combinatorial optimization problem with binary constraints. An unbiased gradient estimation of a stochastic binarization layer is developed for the quantizer unit to generate a bipolar message. As a result, the quantizer and optimizer units of all nodes can be jointly trained via the proposed constrained training algorithm. Then, the real-time computation of the trained DNN units can be implemented in a distributed manner, otherwise not applicable for the DL techniques in \cite{HSun:18,WLee:18b,WLee:18c,WLee:18d,MEisen:18}, since they all require the perfect knowledge of other cells' CSI. Finally, the proposed DL framework is verified from several numerical examples in cognitive multiple access channel (C-MAC) and IFC applications.

This paper is organized as follows: Section \ref{sec:sec2} explains the system model for generic distributed wireless networks. The DL framework for the centralized approach is proposed in Section \ref{sec:sec3}, and the distributed DNN implementation is presented in Section \ref{sec:sec4}. In Section \ref{sec:sec5}, application scenarios of the proposed DL framework are examined and its numerical results are provided. Finally, Section \ref{sec:sec6} conclude the paper.

\textit{Notations:} We employ uppercase boldface letters, lowercase boldface letters, and normal letters for matrices, vectors, and scalar quantities, respectively. A set of $m$-by-$n$ real-valued matrices and a set of length-$m$ bipolar symbol vectors are represented as $\mathbb{R}^{m\times n}$ and $\{-1,+1\}^{m}$, respectively. Let $\mathbb{E}_{X}[\cdot]$ denote the expectation over random variable $X$. In addition, $\mathbf{0}_{m}$ and $\mathbf{1}_{m}$ account for the all zero and all one column vectors of length $m$, respectively.
In the sequel, we use subscripts $C$, $D$, and $Q$ to denote quantities regarding centralized, distributed, and quantization operations, respectively.
Finally, $\nabla_{x}y(x_{0})$ and $\partial_{x}y(x_{0})$ stand for the gradient and the subgradient of $y(x)$ with respect to $x$ evaluated at $x=x_{0}$, respectively.

\section{General Network Setup and Formulation}\label{sec:sec2}
\begin{figure}
\begin{center}
\includegraphics[width=.7\linewidth]{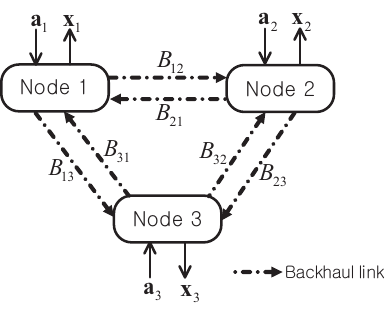}
\end{center}
\caption{Schematic diagram for the distributed network with $N=3$ nodes.}
\label{fig:fig0}
\end{figure}

Fig. \ref{fig:fig0} illustrates a generic distributed network where $N$ nodes desire to minimize the overall system cost in a distributed manner. Node $i$ $(i=1,\ldots,N)$ observes its own local information vector $\mathbf{a}_{i}\in\mathbb{R}^{A_{i}}$ of length $A_{i}$ (e.g., channel coefficient) and computes the optimal state (or the solution) vector $\mathbf{x}_{i}\in\mathbb{R}^{X_{i}}$ of length $X_{i}$ (e.g., the resource allocation strategy). The nodes are interconnected among one another via the backhaul realized by wired or wireless links so that the local observation $\mathbf{a}_{i}$ at node $i$ is shared with neighboring node $j$ ($j\neq i$). The capacity of the backhaul link from node $i$ to node $j$ is limited by $B_{ij}$ in bits/sec/Hz. In practice, a wireless backhaul link is implemented with a reliable control channel orthogonal to wireless access links and dedicated for sharing the control information over the network. Thus, the capacity $B_{ij}$ is fixed and known in advance. The corresponding configuration is formulated in a cost minimization as
\begin{align}
    \text{(P): }& \min_{\mathbf{x}} \quad\mathbb{E}_{\mathbf{a}}[f(\mathbf{a},\mathbf{x})]\nonumber\\
    & \text{subject to} \quad \mathbb{E}_{\mathbf{a}}[g_{k}(\mathbf{a},\mathbf{x})]\leq G_{k},\ k= 1,\ldots,K,\label{eq:C1}\\
    &\qquad \qquad \quad \mathbf{x}_{i}\in\mathcal{X}_{i},\ i=1,\ldots,N,\label{eq:C3}
\end{align}
where $\mathbf{a}\triangleq\{\mathbf{a}_{i}:\forall i\}\in\mathbb{R}^{(\sum_{i=1}^{N}A_{i})}$ and $\mathbf{x}\triangleq\{\mathbf{x}_{i}:\forall i\}\in\mathbb{R}^{(\sum_{i=1}^{N}X_{i})}$ stand for the collection of the local information and the solution of all nodes, respectively. In addition, $f(\mathbf{a},\mathbf{x})$ is a network cost function for given configuration $(\mathbf{a},\mathbf{x})$. 
Here, \eqref{eq:C1} indicates a long-term design constraint expressed by an inequality with the upper bound $G_k$. A class of a long-term design constraint includes the average transmit power budget and the average quality-of-service constraints at users evaluated over fast fading channel coefficients \cite{MMohseni:06,RZhang:09}. 
A convex set $\mathcal{X}_{i}$ in \eqref{eq:C3} denotes the set of feasible assignments $\mathbf{x}_{i}$ for node $i$. The cost function $f(\mathbf{a},\mathbf{x})$ as well as constraint $g_{k}(\mathbf{a},\mathbf{x})$ are assumed to be differentiable but are possibly non-convex.

We desire to optimize the network performance averaged over the observation $\mathbf{a}\in\mathcal{A}$ subject to both long-term and instantaneous regulations. The long-term performance metrics are important in optimizing a wireless network under a fast fading environment where the quality of communication service is typically measured by the average performance, e.g., the ergodic capacity and the average power consumption \cite{EBiglieri:98}. Special cases of (P) have been widely studied in various network configurations such as multi-antenna multi-user systems \cite{MMohseni:06}, CR networks \cite{RZhang:09}, wireless power transfer communications \cite{LLiu:13}, and proactive eavesdropping applications \cite{JXu:17}. Although these studies have successfully addressed the global optimality in their formulations, no generalized framework has not been developed for solving (P) with the guaranteed optimality. Furthermore, existing distributed optimization techniques, such as ADMM and MP algorithms, typically rely on iterative computations which may require sufficiently large capacity backhaul links for exchanging high-precision continuous-valued messages among nodes. This work proposes a DL framework for solving (P) in two individual network configurations. One ideal case is first addressed where backhaul links among the nodes have infinite capacity. The other case corresponds to a more realistic scenario with finite-capacity backhaul links where full cooperation among nodes is not possible.

\subsection{Formulation}
In the infinite-capacity backhaul link case, node $i$ forwards its local observation $\mathbf{a}_{i}$ to other nodes without loss of the information. Thus, the global observation vector $\mathbf{a}=\{\mathbf{a}_{i}:\forall i\in\mathcal{N}\}$ is available to all nodes. With the perfect global information at hand, each node can directly obtain the global network solution $\mathbf{x}$ of (P) from a centralized computation rule $z_{C}(\cdot)$ given by
\begin{align}
    \mathbf{x}=z_{C}(\mathbf{a}), \label{eq:cent}
\end{align}
i.e., the solution $\mathbf{x}$ is a function of global information $\mathbf{a}$ since a node exploits all available information for computing it. The associated optimization (P) is rewritten into (P1) as
\begin{align}
    \text{(P1): }&\min_{z_{C}(\cdot)}\quad \mathbb{E}_{\mathbf{a}}[f(\mathbf{a},z_{C}(\mathbf{a}))]\nonumber\\
    &\text{subject to} \quad \mathbb{E}_{\mathbf{a}}[g_{k}(\mathbf{a},z_{C}(\mathbf{a}))]\leq G_{k},\ k=1,\ldots,K,\nonumber\\
    &\qquad \qquad \quad \mathbf{e}_{i}^{T}z_{C}(\mathbf{a})\in\mathcal{X}_{i},\ i=1,\ldots,N,\nonumber
\end{align}
where $\mathbf{e}_{i}\in\mathbb{R}^{(\sum_{i=1}^{N}X_{i})}$ is a column vector with $X_i$ ones from  $(\sum_{j=1}^{i-1}X_{j}+1)$-th entry to $\sum_{j=1}^{i}X_{j}$-th entry and the remaining entries equal to zero for masking the state associated with node $i$ from the global solution.

In the finite-capacity backhaul case with $B_{ij}<\infty$, node $i$ first discretizes its local measurement $\mathbf{a}_{i}$ to obtain a bipolar quantization $\mathbf{v}_{ij}\in\{-1,+1\}^{B_{ij}}$, which is transferred to node $j~(j\neq i)$. Without loss of generality, it is assumed that $B_{ij}$ is an integer.\footnote{If $B_{ij}$ is not integer, $\lfloor B_{ij}\rfloor$-bit quantization is applied.} The quantization at node $i$ is represented by
\begin{align}
    \mathbf{v}_{i}=z_{Q,i}(\mathbf{a}_{i}),\ i=1,\ldots,N, \label{eq:quant}
\end{align}
where $\mathbf{v}_{i}\triangleq\{\mathbf{v}_{ij}:(i,j), \forall j \neq i\}\in\{-1,+1\}^{L_{i}}$ and $L_{i}\triangleq\sum_{j\neq i}B_{ij}$.
By collecting the information of $\mathbf{a}_{i}$ obtained locally and $\mathbf{v}_{ji}\in\{-1,+1\}^{B_{ji}}$ transferred from neighbors, node $i$ calculates its state  $\mathbf{x}_{i}$ via a certain computation rule $z_{D,i}(\cdot)$ as
\begin{align}
    \mathbf{x}_{i}=z_{D,i}(\mathbf{c}_{i}),\ i=1,\ldots,N, \label{eq:dist}
\end{align}
where $\mathbf{c}_{i}$ is the concatenation of $\mathbf{a}_{i}$ and $\{\mathbf{v}_{ji}:(j,i), \forall j\neq i\}$. Since \eqref{eq:dist} only requires the knowledge of the local observation and the distributed computation, it is referred to as a distributed approach. The distributed realization consists in jointly designing bipolar quantization \eqref{eq:quant} and distributed optimization \eqref{eq:dist} for all nodes. The optimization (P) can be refined as
\begin{align}
    \text{(P2): }&\min_{\{z_{D,i}(\cdot),z_{Q,i}(\cdot)\}}\quad\mathbb{E}_{\mathbf{a}}[f(\mathbf{a},\{z_{D,i}(\mathbf{c}_{i}):\forall i\})]\nonumber\\
    &\text{subject to} \quad \mathbb{E}_{\mathbf{a}}[g_{k}(\mathbf{a},\{z_{D,i}(\mathbf{c}_{i}):\forall i\})]\leq G_k,\nonumber\\
    & \qquad \qquad \quad k=1,\ldots,K,\nonumber\\
    &\qquad \qquad \quad z_{D,i}(\mathbf{c}_{i})\in\mathcal{X}_{i},\ i=1,\ldots,N,\nonumber\\
    &\qquad \qquad \quad \mathbf{v}_{i}=z_{Q,i}(\mathbf{a}_{i})\in\{-1,+1\}^{L_{i}},\nonumber\\
    &\qquad \qquad \quad i=1,\ldots,N.\label{eq:binary}
\end{align}
Note that the binary constraint in \eqref{eq:binary} incurs a computationally demanding computation of search for efficient quantization $\mathbf{v}_{i}$ associated with sample $\mathbf{a}_{i}$. Thus, solving (P2) by traditional optimization techniques is not straightforward in general.

\subsection{Learning to Optimize}
To address constrained optimization problems (P1) and (P2), a DL framework that identifies unknown functions in \eqref{eq:cent}-\eqref{eq:dist} is developed. The basics of a feedforward DNN with fully-connected layers are introduced briefly. Let $\mathbf{u}_{R+1}=\phi(\mathbf{u}_{0};\theta)$ be a DNN with $R$ hidden layers that maps input vector $\mathbf{u}_{0}\in\mathbb{R}^{U_{0}}$ to  output $\mathbf{u}_{R+1}\in\mathbb{R}^{U_{R+1}}$ with a set of parameters $\theta$. Subsequently, $\mathbf{u}_{r}\in\mathbb{R}^{U_{r}}$ denotes $U_{r}$-dimensional output of layer $r$, for $r=1, \ldots, R, R+1$, and can be expressed by
\begin{align}
    \mathbf{u}_{r}=\zeta_{r}(\mathbf{W}_{r}\mathbf{u}_{r-1}+\mathbf{b}_{r}),\ r=1,\ldots,R,R+1, \label{eq:dnn}
\end{align}
where $\mathbf{W}_{r}\in\mathbb{R}^{U_{r}\times U_{r-1}}$ and $\mathbf{b}_{r}\in\mathbb{R}^{U_{r}}$ account for the weight matrix and the bias vector at layer~$r$, respectively. An element-wise function $\zeta_{r}(\cdot)$ is the activation at layer $r$. The set of parameters $\theta$ is defined by $\theta\triangleq\{\mathbf{W}_{r},\mathbf{b}_{r}:\forall r\}$. The input-output relationship $\mathbf{u}_{R+1}=\phi(\mathbf{u}_{0};\theta)$ is specified by the successive evaluation of layers in \eqref{eq:dnn}. The training step determines the DNN parameter $\theta$ such that the cost function of the DNN is minimized. The activation of the DNN computation in \eqref{eq:dnn} typically involves non-convex operations, and thus a closed-form expression is not available for the optimal parameter set $\theta$. State-of-the-art DL libraries employ a gradient decent (GD) method and its variants for stochastic optimization \cite{LeCun:15}. The details of the proposed DNN construction and the corresponding training methods are described in the following sections.

DL techniques have been intensively investigated to solve optimization challenges in wireless networks such as power control and scheduling problems over the IFC \cite{WLee:18a,PKerret:18,HSun:18}, CR networks \cite{WLee:18b}, and device-to-device communications \cite{WLee:18c,WLee:18d}. However, the long-term constraint \eqref{eq:C1} has not been properly considered in the DNN training. In addition, the existing techniques are mostly intended for a centralized solution, which is not necessarily feasible in practice. Several heuristics for the distributed realization have been provided for the IFC in \cite{WLee:18a} and \cite{PKerret:18}. Those techniques, however, lack an optimized quantization, which is a crucial feature of the distributed network for communication cost saving. It is still not straightforward to obtain efficient solutions of (P1) and (P2) with the existing DL techniques. In the following sections, a constrained training algorithm is first developed for solving (P1) with the long-term constraints based on the primal-dual method \cite{Boyd:03,Boyd:04}. Subsequently, a binarization operation addresses the binary constraint \eqref{eq:binary} in the distributed approach.

\section{Centralized Approach}\label{sec:sec3}
This section develops a DL framework that solves the constrained optimization in (P1) with the DNN that evaluates centralized computation \eqref{eq:cent}. To this end, unknown function $\mathbf{x}=z_{C}(\mathbf{a})$ is replaced by a DNN $\phi_{C}(\mathbf{a};\mathbf{\theta}_{C})$ as
\begin{align}
    \mathbf{x}=\phi_{C}(\mathbf{a};\mathbf{\theta}_{C}).\label{eq:cent_dnn}
\end{align}
The following proposition assesses the quality of an approximation for $\mathbf{x}=z_{C}(\mathbf{a})$ by the DNN in \eqref{eq:cent_dnn} based on the universal approximation theorem \cite{KHornik:89}, which investigates the existence of the DNN $\phi_{C}(\mathbf{a};\mathbf{\theta}_{C})$ with an arbitrary worst-case approximation error over the training set $\mathbf{a}\in\mathcal{A}$.
\begin{prop}\label{lem:lem2}
Let $z_{C}(\mathbf{a})$ be a continuous function defined over a compact set $\in\mathcal{A}=\{\mathbf{a}\}$. Suppose that the DNN $\phi_{C}(\mathbf{a};\theta_{C})$ with $R_{C}$ hidden layers are constructed by sigmoid activations. Then, for any $\varepsilon>0$, there exists a sufficiently large number $R_{C}>0$ such that
\begin{align}
    \sup_{\mathbf{a}\in\mathcal{A}}\|z_{C}(\mathbf{a})-\phi_{C}(\mathbf{a};\theta_{C})\|\leq\varepsilon.\label{eq:unv}
\end{align}
\end{prop}
\begin{IEEEproof}
Please refer to \cite{KHornik:89} and \cite[Theorem 1]{HSun:18}.
\end{IEEEproof}
Proposition \ref{lem:lem2}, which is based on the universal approximation theorem \cite{KHornik:89}, implies that for a given set $\mathcal{A}$, there exist a set parameter $\theta_{C}$ such that the associated DNN of the structure in \eqref{eq:cent_dnn} can approximate any continuous function $z_{C}(\mathbf{a})$ with arbitrary small positive error $\varepsilon$. This holds both for supervised and unsupervised learning problems, regardless of the convexity of cost and constraint functions. Therefore, the unknown optimal computation process $z_{C}^{\star}(\mathbf{a})$ for the non-convex problem (P1) is successfully characterized by a well-designed DNN $\phi_{C}(\mathbf{a};\theta_{C})$. Proposition \ref{lem:lem2} only states the existence of $\theta_{C}$ satisfying (\ref{eq:unv}) rather than a specific identification method. To determine $\theta_{C}$, the DNN is trained such that it yields an efficient solution for (P1). 

\begin{rem}
Recently, it is revealed in \cite{ZLu:17} that, for any $\varepsilon>0$ and Lebesgue-integrable function $z_{C}(\mathbf{a})$, i.e., $\int_{\mathcal{A}}|z_{C}(\mathbf{a})| \mathrm{d}\mathbf{a}<\infty$, there exists a fully-connected DNN $\phi_{C}(\mathbf{a};\theta_{C})$ with rectified linear unit (ReLU) activations that satisfy 
$\int_{\mathcal{A}}|z_{C}(\mathbf{a})-\phi_{C}(\mathbf{a};\theta_{C})| \mathrm{d}\mathbf{a}<\varepsilon.$ 
The width of $\phi_{C}(\mathbf{a};\theta_{C})$, i.e., the maximum number of nodes in hidden layers is bounded by $\sum_{i=1}^{N}A_{i}+4$, and the upperbound for the number of hidden layers is given by $4\sum_{i=1}^{N}A_{i}+1$. From (\ref{eq:quant}), it can be concluded that the DNN can act as an universal approximator for a non-continuous and Lebesgue-integrable function, e.g., an indicator function. This allows a DNN to learn discrete mappings in wireless resource management including the quantization operation in (\ref{eq:quant}).
\end{rem}

Next, we construct the DNN \eqref{eq:cent_dnn} for efficiently solving (P1) under the deterministic constraint $\mathbf{x}_{i}\in\mathcal{X}_{i}$. Suppose that the DNN $\phi_{C}(\mathbf{a};\theta_{C})$ consists of $R_{C}$ hidden layers and parameter set $\theta_{C}\triangleq\{\mathbf{W}_{C,r},\mathbf{b}_{C,r}:\forall r=1,\ldots,R_{C}+1\}$. The dimension, the activation functions, and the number of hidden layers of the DNN are hyper parameters to be optimized via a validation. By contrast, the dimension of the output layer is fixed to $\sum_{i=1}^{N}X_{i}$ for obtaining the solution $\mathbf{x}\in\mathbb{R}^{(\sum_{i=1}^{N}X_{i})}$. To satisfy the masked constraint $\mathbf{x}_{i}=\mathbf{e}_{i}^{T}\phi_{C}(\mathbf{a};\mathbf{\theta}_{C})\in\mathcal{X}_{i}$, the activation $\zeta_{R_{C}+1}(\cdot)$ is set to a projection operation $\Pi_{\mathcal{X}_{i}}(\mathbf{u})\triangleq\arg\min_{\mathbf{x}_{i}\in\mathcal{X}_{i}}\|\mathbf{u}-\mathbf{x}\|^{2}$ onto the convex set $\mathcal{X}_{i}$
\begin{align}
    \mathbf{x}_{i}=\Pi_{\mathcal{X}_{i}}(\mathbf{e}_{i}^{T}(\mathbf{W}_{C,R_{C}+1}\mathbf{u}_{C,R_{C}}+\mathbf{b}_{C,R_{C}+1})),\ i=1,\ldots,N, \label{eq:proj}
\end{align}
where $\mathbf{u}_{C,r}\in\mathbb{R}^{U_{C,r}}$ is the output vector of hidden layer $r$ of the DNN $\phi_{C}(\mathbf{a};\mathbf{\theta}_{C})$.
Since a convex projection can be realized with linear operations, the gradient of the activation in \eqref{eq:proj} that solves the convex projection problem is well-defined \cite{BAmos:17}, and the backpropagation works well with this activation. The feasibility of such a projection layer has been verified in \cite{HLee:18a}. 
Plugging \eqref{eq:cent_dnn} and \eqref{eq:proj} into (P1) leads to
\begin{align}
    \text{(P1.1): }&\min_{\theta_{C}}\quad \mathbb{E}_{\mathbf{a}}[f(\mathbf{a},\phi_{C}(\mathbf{a};\mathbf{\theta}_{C}))]\nonumber\\
    &\text{subject to} \quad \mathbb{E}_{\mathbf{a}}[g_{k}(\mathbf{a},\phi_{C}(\mathbf{a};\mathbf{\theta}_{C}))]\leq G_{k},\nonumber\\
    &\qquad \qquad \quad k=1,\ldots,K,\label{eq:P11C1}
\end{align}
where the constraint $\mathbf{x}_{i}\in\mathcal{X}_{i}$ is lifted by the convex projection activation \eqref{eq:proj}. In (P1.1), the optimization variable is now characterized by the DNN parameter set $\theta_{C}$, which can be more efficiently handled by state-of-the-art DL libraries including TensorFlow as compared to the direct determination of unknown function $z_{C}(\mathbf{a})$. One major challenge in (P1.1) features the integration of constraints in \eqref{eq:P11C1} into the training task of the DNN parameter set $\theta_{C}$. It is not straightforward to address this issue by existing DL training algorithms which originally focus on unconstrained training applications. Recent DL applications in wireless communications mostly resort to a penalizing technique which transforms to an unconstrained training task by augmenting an appropritate regularization \cite{WLee:18b,WLee:18c,HLee:18b}. However, the feasibility of the trained DNN is not ensured analytically with an arbitrary choice of a hyperparameter on the regularization. Therefore, the hyperparameter is carefully adjusted via the evaluation of the validation performance during the DL training. The hyperparameter optimization is typically carried out through a trial-and-error search which often incurs computationally demanding calculations.

To overcome this issue, a Lagrange duality method \cite{Boyd:04} is applied to include the constraints of (P1.1) to the training step. The non-convexity of cost and constraint functions with respect to $\theta_{C}$ may result in a positive duality gap. Nevertheless, the following proposition verifies that (P1.1) fulfills the \textit{time-sharing} condition \cite{WYu:06}, which guarantees the zero duality gap.
\begin{prop}\label{lem:lem1}
Let $G_{k}^{\min}\triangleq\min_{\theta_{C}}\mathbb{E}_{\mathbf{a}}[g_{k}(\mathbf{a},\phi_{C}(\mathbf{a};\theta_{C}))]$ and $G_{k}^{\max}\triangleq\max_{\theta_{C}}\mathbb{E}_{\mathbf{a}}[g_{k}(\mathbf{a},\phi_{C}(\mathbf{a};\theta_{C}))]$ be the minimum and the maximum achievable upper bound $G_{k}$ of the constraint \eqref{eq:P11C1}, respectively. Suppose that there exists an arbitrary small number $\delta\in(0,G_{k}^{\max}-G_{k}^{\min}]$. Consider two distinct achievable upper bounds $\tilde{G}_{k}$ and $\hat{G}_{k}$ such that $\tilde{G}_{k},\hat{G}_{k}\in[G_{k}^{\min}+\delta,G_{k}^{\max}],~\forall k$. 
We define $\tilde{\theta}_{C}$ and $\hat{\theta}_{C}$ as the optimal solution of (P1.1) with the upper bound $\{G_{k}:\forall k\}$ being replaced by $G_{k}=\tilde{G}_{k}$ and $G_{k}=\hat{G}_{k}$, $\forall k$, respectively.
Then, for arbitrary constant $\tau\in[0,1]$, we can find a feasible DNN parameter set $\theta_{C}$ which satisfies
\begin{align}
    \mathbb{E}_{\mathbf{a}}[f(\mathbf{a},\phi_{C}(\mathbf{a};\theta_{C}))]&\leq \tau\mathbb{E}_{\mathbf{a}}[f(\mathbf{a},\phi_{C}(\mathbf{a};\tilde{\theta}_{C}))]\nonumber\\
         &~~+ (1-\tau)\mathbb{E}_{\mathbf{a}}[f(\mathbf{a},\phi_{C}(\mathbf{a};\hat{\theta}_{C}))],\label{eq:time1}\\
    \mathbb{E}_{\mathbf{a}}[g_{k}(\mathbf{a},\phi_{C}(\mathbf{a};\theta_{C}))]&\leq \tau \tilde{G}_{k} + (1-\tau)\hat{G}_{k},\ k\in [1:K].\label{eq:time2}
\end{align}
\end{prop}
\begin{IEEEproof}
The proof proceeds similarly as in \cite[Theorem 2]{WYu:06}. Let $\theta_{C}$ be a parameter corresponding to the time-sharing point between $\tilde{\theta}_{C}$ and $\hat{\theta}_{C}$, i.e., by setting $\theta_{C}=\tilde{\theta}_{C}$ for fraction $\tau$ and $\theta_{C}=\hat{\theta}_{C}$ for the remaining fraction $1-\tau$\cite{WYu:06}. It is obvious that the equality in (\ref{eq:time1}) holds for such a configuration $\theta_{C}$ and the constraints in (\ref{eq:time2}) are also satisfied. As a result, it is concluded that (P1.1) fulfils the time-sharing condition.
\end{IEEEproof}
Proposition \ref{lem:lem1} addresses the existence of a feasible $\theta_{C}$ satisfying the time sharing condition in \eqref{eq:time1} and \eqref{eq:time2}, for $\tilde{G}_{k},\hat{G}_{k}\in[G_{k}^{\min}+\delta,G_{k}^{\max}]$, $\forall k$.
Let $F^{\star}(\{G_{k}\})$ be the optimal value of (P1.1) with a nontrivial upper bound $G_{k}\in[G_{k}^{\min}+\delta,G_{k}^{\max}]$, $\forall k$. Then, Slater's condition holds for (P1.1), i.e., for any $G_{k}\in[G_{k}^{\min}+\delta,G_{k}^{\max}]$ with arbitrary small $\delta\in(0,G_{k}^{\max}-G_{k}^{\min}]$, there is a strictly feasible $\theta_{C}$ such that $\mathbb{E}_{\mathbf{a}}[g_{k}(\mathbf{a},\phi_{C}(\mathbf{a};\theta_{C}))]<G_{k}$, $\forall k$. The analysis in \cite{WYu:06} and \cite{TTerlaky:08}, which combines the time sharing property in Proposition \ref{lem:lem1} with Slater's condition, indicates that the optimal objective value $F^{\star}(\{G_{k}\})$ is convex in nontrivial regime $G_{k}\in[G_{k}^{\min}+\delta,G_{k}^{\max}]$ and ensures the strong duality for the non-convex problem (P1.1) \cite{Boyd:04}. Therefore, the Lagrange duality method can be employed to solve (P1.1).
The Lagrangian of (P1.1) is formulated as
\begin{align}
    \mathcal{L}(\theta_{C},\boldsymbol{\lambda})=&\mathbb{E}_{\mathbf{a}}[f(\mathbf{a},\phi_{C}(\mathbf{a};\theta_{C}))]\nonumber\\
        &+\sum_{k=1}^{K}\lambda_{k}(\mathbb{E}_{\mathbf{a}}[g_{k}(\mathbf{a},\phi_{C}(\mathbf{a};\theta_{C}))]-G_{k}),\nonumber
\end{align}
where a nonnegative $\lambda_{k}$ corresponds to the dual variable associated with each constraint in \eqref{eq:P11C1}, and their collection is denoted by $\boldsymbol{\lambda}\triangleq\{\lambda_{k}:\forall k\}\in\mathbb{R}^{K}$. The dual function $\mathcal{G}(\boldsymbol{\lambda})$ is then defined by
\begin{align}
    \mathcal{G}(\boldsymbol{\lambda})=\min_{\theta_{C}}\mathcal{L}(\theta_{C},\boldsymbol{\lambda}), \nonumber
\end{align}
and the dual problem is written as
\begin{align}
    &\max_{\boldsymbol{\lambda}} \quad \mathcal{G}(\boldsymbol{\lambda}) \nonumber\\
    &\text{subject to} \quad \lambda_{k}\geq 0,\ k=1,\ldots,K. \label{eq:dual_prob}
\end{align}
To solve the problem in \eqref{eq:dual_prob}, the primal-dual method is employed to perform iterative updates between primal variable $\theta_{C}$ and dual variable $\boldsymbol{\lambda}$ \cite{Boyd:03}.\footnote{A recent work studies a primal-dual method for the constrained training technique based on independent analysis for the strong duality \cite{MEisen:18}. However, the algorithm in \cite{MEisen:18} requires additional gradient computation to update auxiliary variables.} For the convenience of the representation, let $x^{[t]}$ denote the value of $x$ evaluated at the $t$-th iteration of the update. The primal update for $\theta_{C}^{[t]}$ can be calculated by the GD method as
\begin{align}
    \theta_{C}^{[t]}&=\theta_{C}^{[t-1]}-\eta\nabla_{\theta_{C}}\mathcal{L}(\theta_{C}^{[t-1]},\boldsymbol{\lambda}^{[t-1]})\nonumber\\
        &=\theta_{C}^{[t-1]}-\eta\bigg(\mathbb{E}_{\mathbf{a}}[\nabla_{\theta_{C}}f(\mathbf{a},\phi_{C}(\mathbf{a};\theta_{C}^{[t-1]}))]\nonumber\\
        &~~+\sum_{k=1}^{K}\lambda_{k}^{[t-1]}(\mathbb{E}_{\mathbf{a}}[\nabla_{\theta_{C}}g_{k}(\mathbf{a},\phi_{C}(\mathbf{a};\theta_{C}^{[t]}))]-G_{k})\bigg)
    , \label{eq:p_up}
\end{align}
where a positive $\eta$ represents the learning rate. The gradient $\nabla_{\theta_{C}}\psi(\mathbf{a},\phi_{C}(\mathbf{a};\theta_{C}))$ for a function in $\psi(\cdot)\in\{f(\cdot),g_{k}(\cdot):\forall k\}$ is given by the chain rule as
\begin{align}
    \nabla_{\theta_{C}}\psi(\mathbf{a},\phi_{C}(\mathbf{a};\theta_{C}))=&\nabla_{\phi_{C}(\mathbf{a};\theta_{C})}\psi(\mathbf{a},\phi_{C}(\mathbf{a};\theta_{C}))\nonumber\\
    &\cdot\nabla_{\theta_{C}}\phi_{C}(\mathbf{a};\theta_{C}).\nonumber
\end{align}
The dual update for $\lambda_{k}^{[t]}$ of the dual problem \eqref{eq:dual_prob} is determined by utilizing the projected subgradient method \cite{Boyd:03} as
\begin{align}
    \lambda_{k}^{[t]}&=\Big({\lambda_{k}}^{[t-1]}+\eta\partial_{\lambda_{k}}\mathcal{G}(\theta_{C}^{[t-1]},\boldsymbol{\lambda}^{[t-1]})\Big)_{+}\nonumber\\
        &=\Big({\lambda_{k}}^{[t-1]}+\eta(\mathbb{E}_{\mathbf{a}}[g_{k}(\mathbf{a},\phi_{C}(\mathbf{a};\theta_{C}^{[t-1]}))]-G_{k})\Big)_{+},
        \label{eq:d_up}
\end{align}
Proposition 1 leads to a DNN structure $\phi_{C}(\mathbf{a};\theta_{C})$ which can compute the dual variable via the update rule in (\ref{eq:d_up}). Based on \eqref{eq:p_up} and \eqref{eq:d_up}, the DNN parameter $\theta_{C}^{[t]}$ and the dual solution $\boldsymbol{\lambda}^{[t]}$ can be jointly optimized by a single GD training computation.
To implement the expectations \eqref{eq:p_up} and \eqref{eq:d_up} in practice, we adopt the mini-batch SGD algorithm, which is a powerful stochastic optimization tool for DL \cite{LeCun:15}, to exploit parallel computing capability of general-purpose graphical processing units (GPUs). At each iteration, mini-batch set $\mathcal{S}\subset\mathcal{A}$ of size $S$ is either sampled from training set $\mathcal{A}$ or generated from the probability distribution of global observation vector $\mathbf{a}$, if available. The update rules in \eqref{eq:p_up} and \eqref{eq:d_up} are replaced by
\begin{align}
    \theta_{C}^{[t]}&=\theta_{C}^{[t-1]}\nonumber\\
    &-\eta\bigg(\frac{1}{S}\sum_{\mathbf{a}\in\mathcal{S}}\nabla_{\theta_{C}}f(\mathbf{a},\phi_{C}(\mathbf{a};\theta_{C}^{[t-1]}))\nonumber\\
        &+\sum_{k=1}^{K}\lambda_{k}^{[t-1]}\Big(\frac{1}{S}\sum_{\mathbf{a}\in\mathcal{S}}\nabla_{\theta_{C}}g_{k}(\mathbf{a},\phi_{C}(\mathbf{a};\theta_{C}^{[t-1]}))-G_{k}\Big)\bigg),\label{eq:p_up2}\\
    \lambda_{k}^{[t]}&=\bigg({\lambda_{k}}^{[t-1]}+\eta\Big(\frac{1}{S}\sum_{\mathbf{a}\in\mathcal{S}}g_{k}(\mathbf{a},\phi_{C}(\mathbf{a};\theta_{C}^{[t-1]}))-G_{k}\Big)\bigg)_{+}.\label{eq:d_up2}
\end{align}
Note in (\ref{eq:p_up2}) and (\ref{eq:d_up2}) that average cost and constraint functions are approximated with a sample mean over mini-batch set $\mathcal{S}$. The number of mini-batch samples $S$ is chosen to a sufficiently large number for accurate primal and dual updates. Its impact is discussed via numerical results in Section V.

\begin{algorithm}
\caption{Proposed constrained training algorithm}
\begin{algorithmic}
    \STATE Initialize $t=0$, $\theta^{[0]}$ and $\boldsymbol{\lambda}^{[0]}$.
    \REPEAT{}
        \STATE Set $t\leftarrow t+1$.
        \STATE Sample a mini-batch set $\mathcal{S}\subset\mathcal{A}$.
        \STATE Update $\theta_{C}^{[t]}$ and $\boldsymbol{\lambda}^{[t]}$ from \eqref{eq:p_up2} and \eqref{eq:d_up2}, respectively.
    \UNTIL convergence
\end{algorithmic}
\end{algorithm}

Algorithm 1 summarizes the overall constrained training algorithm. Unlike the supervised learning method in \cite{HSun:18} which trains the DNN to memorize given labels obtained by solving the optimization explicitly, the proposed algorithm employs an unsupervised learning strategy where the DNN learns how to identify an efficient solution to (P1.1) without any prior knowledge. In addition to the DNN parameter $\theta_{C}$, the dual variables in \eqref{eq:dual_prob} are optimized together, and thus the primal feasibility for \eqref{eq:P11C1} is always guaranteed upon convergence \cite{Boyd:03,Boyd:04}. Once the DNN $\phi_{C}(\mathbf{a};\theta_{C})$ is trained from Algorithm 1, the parameter $\theta_{C}$ is configured at nodes. Subsequently, node $i$ computes its solution $\mathbf{x}_{i}$ using $\mathbf{x}_{i}=\mathbf{e}_{i}^{T}\phi_{C}(\mathbf{a};\theta_{C})$ with the collection of local observations $\{\mathbf{a}_{j}\}$ received from other node $j~(j\neq i)$. The performance of the trained DNN is subsequently evaluated with testing samples which are unseen during the training.

\section{Distributed Approach}\label{sec:sec4}


\begin{figure}
\centering
    \subfigure[Optimizer and quantizer DNNs]{
        \includegraphics[width=.8\linewidth]{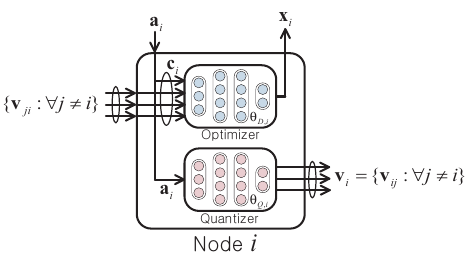}\label{fig:fig1}
    }
    \subfigure[Quantizer DNN with the stochastic binarization layer.]{
        \includegraphics[width=.8\linewidth]{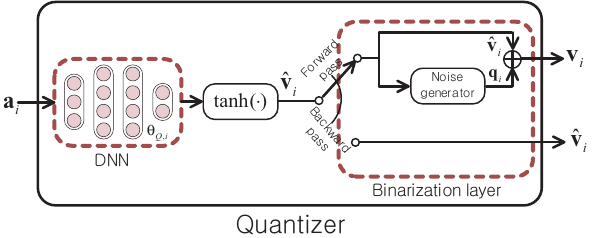}\label{fig:fig2}
    }
    \caption{Structure of node $i$ in the proposed distributed approach.}
    \label{fig:fig12}
\end{figure}

The DL method proposed in Section \ref{sec:sec3} relies on the perfect knowledge of observations from other nodes, which is realized by the infinite-capacity backhaul links among nodes. This section presents a DL based distributed approach for (P2) by jointly optimizing the quantization \eqref{eq:quant} and the distributed optimization \eqref{eq:dist}. Similar to the centralized approach, DNNs produce $\mathbf{v}_{i}\triangleq\{\mathbf{v}_{ij}:(i,j)~\forall j\neq i\}=\phi_{Q,i}(\mathbf{a}_{i};\theta_{Q,i})\in\mathbb{R}^{L_{i}}$ and $\mathbf{x}_{i}=\phi_{D,i}(\mathbf{c}_{i};\theta_{D,i})\in\mathbb{R}^{X_{i}}$ that approximate \eqref{eq:quant} and \eqref{eq:dist} with $\mathbf{c}_{i}\triangleq\{\mathbf{a}_{i},\mathbf{v}_{ji}:(j,i)~\forall j\neq i\}$, respectively. The universal approximation property in Proposition \ref{lem:lem2} secures the accurate description of the optimal solution $z_{D,i}^{\star}(\mathbf{c}_{i})$ and $z_{Q,i}^{\star}(\mathbf{a}_{i})$ for (P2) with DNNs $\phi_{D,i}(\mathbf{c}_{i};\theta_{D,i})$ and $\phi_{Q,i}(\mathbf{a}_{i};\theta_{Q,i})$, respectively. As illustrated in Fig. \ref{fig:fig1}, a node is equipped with two different DNNs, i.e., the quantizer unit $\phi_{Q,i}(\mathbf{a}_{i};\theta_{Q,i})$ and the distributed optimizer unit $\phi_{D,i}(\mathbf{c}_{i};\theta_{D,i})$. This results in the joint training of total of $2N$ DNNs for distributed implementation. For notational convenience, the collection of the optimizer units from all nodes is denoted by $\phi_{D}(\mathbf{c};\theta_{D})\triangleq\{\phi_{D,i}(\mathbf{c}_{i};\theta_{D,i}):\forall i\}$, where $\mathbf{c}\triangleq\{\mathbf{c}_{i}:\forall i\}$ and $\theta_{D}\triangleq\{\theta_{D,i}:\forall i\}$. Subsequently, the distributed optimization task (P2) can be transformed as
\begin{align}
    \text{(P2.1): }&\min_{\{\theta_{D,i},\theta_{Q,i}\}}\quad \mathbb{E}_{\mathbf{a}}[f(\mathbf{a},\phi_{D}(\mathbf{c};\theta_{D}))]\nonumber\\
    &\text{subject to} \quad \mathbb{E}_{\mathbf{a}}[g_{k}(\mathbf{a},\phi_{D}(\mathbf{c};\theta_{D}))]\leq G_{k},\nonumber\\
    &\qquad \qquad \quad k=1,\ldots,K,\label{eq:P21C0}\\
    &\qquad \qquad \quad \mathbf{v}_{i}=\phi_{Q,i}(\mathbf{a}_{i};\theta_{Q,i})\in\{-1,+1\}^{L_{i}},\nonumber\\
    &\qquad \qquad \quad i=1,\ldots,N,\label{eq:P21C1}
\end{align}
where the deterministic convex constraint $\phi_{D,i}(\mathbf{c}_{i};\theta_{D,i})\in\mathcal{X}_{i}$ is lifted by the convex projection activation in \eqref{eq:proj} at the output layer of each optimizer unit $\phi_{D,i}(\mathbf{c}_{i};\theta_{D,i})$. Note that it is very difficult to enforce the non-convex binary constraint in \eqref{eq:P21C1} with a projection activation, such as signum function $\text{sign}(x)$, since a well-known \textit{vanishing gradient} issue occurs in the DNN training step \cite{LeCun:15}. To see this more carefully, $\text{sign}(x)$ can be chosen for the activation at the output layer of the quantizer DNN~$\phi_{Q,i}(\mathbf{a}_{i};\theta_{Q,i})$ to yield binary output vector $\mathbf{v}_{i}$. However, it has a null gradient for entire input range. The quantizer DNN parameter $\theta_{Q,i}$ is not trained at all by the SGD algorithm. Thus, such a deterministic binarization activation does not allow to train the DNNs effectively by using existing DL libraries.

\subsection{Stochastic binarization layer}


Fig. \ref{fig:fig2} presents the proposed stochastic binarization layer that enables a SGD based training algorithm while satisfying the binary constraint in \eqref{eq:P21C1}. For binarization, hyperbolic tangent function $\text{tanh}(x)\triangleq\frac{e^{2x}-1}{e^{2x}+1}$ is the activation at the output layer of the quantizer DNN $\phi_{Q,i}(\mathbf{a}_{i};\theta_{Q,i})$ to restrict the output $\hat{\mathbf{v}}_{i}\in\mathbb{R}^{L_{i}}$ within a $L_{i}$-dimensional square box, i.e., $\hat{\mathbf{v}}_{i} \in [-1,1]^{L_{i}}$. Since the output $\hat{\mathbf{v}}_{i}$ is still a continuous-valued vector, the final bipolar output $\mathbf{v}_{i}\in\{-1,+1\}^{L_{i}}$ is produced via a stochastic operation on continuous-valued vector $\hat{\mathbf{v}}_{i}$ as
\begin{align}
    \mathbf{v}_{i}=\hat{\mathbf{v}}_{i}+\mathbf{q}_{i},\label{eq:sto_bin}
\end{align}
where the $l$-th element $q_{i,l}$ of $\mathbf{q}_{i}\triangleq[q_{i,1},\ldots,q_{i,L_{i}}]^{T}\in\mathbb{R}^{L_{i}}$ is a random variable given by
\begin{align}\label{eq:q}
    q_{i,l}=\begin{cases}
    1-\hat{v}_{i,l}, & \text{with probability}\ \beta(\hat{v}_{i,l}),\\
    -1-\hat{v}_{i,l}, & \text{with probability}\ 1-\beta(\hat{v}_{i,l}),
  \end{cases}
\end{align}
where $\hat{v}_{i,l}$ corresponds to the $l$-th element of $\hat{\mathbf{v}}_{i}\triangleq[\hat{v}_{i,1},\ldots,\hat{v}_{i,L_{i}}]^{T}$ and $\beta(x)\in[0,1]$ determines the probability that variable $x$ is mapped to one at the output. Note that \eqref{eq:sto_bin} and \eqref{eq:q} guarantee a bipolar output $\mathbf{v}_{i}\in\{-1,+1\}^{L_{i}}$. The stochastic operation in \eqref{eq:sto_bin} can be regarded as a binarization operation of the unquantized vector $\hat{\mathbf{v}}_{i}$, and $\mathbf{q}_{i}$ corresponds to the quantization noise. In \eqref{eq:q}, any function of nonzero gradient can be chosen for a candidate of $\beta(x)$. This work employs an affine function $\beta(x)=\frac{1+x}{2}$ for $x\in[-1,1]$ leading to the mean of noise $\mathbf{q}_{i}$ equal to zero as
\begin{align}
    \mathbb{E}_{\mathbf{q}_{i}}[\mathbf{q}_{i}]
    =&\hat{\mathbf{v}}_{i}+(\mathbf{1}_{L_{i}}-\hat{\mathbf{v}}_{i})\odot\frac{1+\hat{\mathbf{v}}_{i}}{2}\nonumber\\
        &+(-\mathbf{1}_{L_{i}}-\hat{\mathbf{v}}_{i})\odot\frac{1-\hat{\mathbf{v}}_{i}}{2}
    =\mathbf{0}_{L_{i}},\label{eq:q_mean}
\end{align}
where $\odot$ denotes the element-wise multiplication. The vanishing mean in \eqref{eq:q_mean} facilitates the back propagation implementation of the stochastic binarization layer \eqref{eq:sto_bin}.

It still remains unaddressed to find the gradient of the stochastic binarization in \eqref{eq:sto_bin}, i.e., $\nabla_{\theta_{Q,i}}\mathbf{v}_{i}$. No closed-form formula is available due to stochastic noise $\mathbf{q}_{i}$. To resolve this, a gradient estimation idea \cite{YBengio:13,Raiko:15} is adopted such that the gradient of the stochastic operation in \eqref{eq:sto_bin} is approximated by its expectation averaged over the quantization noise distribution, i.e., $\nabla_{\theta_{Q,i}}\mathbf{v}_{i}\simeq\nabla_{\theta_{Q,i}}\mathbb{E}_{\mathbf{q}_{i}}[\mathbf{v}_{i}|\hat{\mathbf{v}}_{i}]$. The gradient is derived from \eqref{eq:q_mean} as
\begin{align}
    \nabla_{\theta_{Q,i}}\mathbf{v}_{i}\simeq&\nabla_{\theta_{Q,i}}\mathbb{E}_{\mathbf{q}_{i}}[\mathbf{v}_{i}|\hat{\mathbf{v}}_{i}]
        =\nabla_{\theta_{Q,i}}\mathbb{E}_{\mathbf{q}_{i}}[(\hat{\mathbf{v}}_{i}+\mathbf{q}_{i})|\hat{\mathbf{v}}_{i}]\nonumber\\
        =&\nabla_{\theta_{Q,i}}\hat{\mathbf{v}}_{i},\label{eq:est}
\end{align}
Here, the zero-mean quantization noise replaces the gradient of $\mathbf{v}_{i}$ with a deterministic gradient of the unquantized vector $\hat{\mathbf{v}}_{i}$, which corresponds to the output of the hyperbolic tangent activation with a non-vanishing gradient. In addition, (\ref{eq:est}) reveals that the bipolar vector $\mathbf{v}_{i}$ is an unbiased estimator of the unquantized vector $\hat{\mathbf{v}}_{i}$, i.e., $\mathbb{E}_{\mathbf{q}_{i}}[\mathbf{v}_{i}|\hat{\mathbf{v}}_{i}]=\hat{\mathbf{v}}_{i}$. Note from \eqref{eq:sto_bin} and \eqref{eq:est} that, during the training step, the operation of the stochastic binarization layer is different for the forward pass and the backward pass. In the forward pass where the DNN calculates the output from the input and, subsequently, the cost function, the binarization layer of the quantizer unit passes actual quantization vector $\mathbf{v}_{i}$ in \eqref{eq:sto_bin} to the optimizer unit, so that the exact values of the cost and constraint functions in (P2.1) are evaluated with the binary constraint \eqref{eq:P21C1}. On the other hand, in the backward pass, the unquantized value $\hat{\mathbf{v}}_{i}$ of the binary vector $\mathbf{v}_{i}$ is delivered from the output layer to the input layer of $\phi_{Q,i}(\mathbf{a};\theta_{Q,i})$, since the gradient of $\mathbf{v}_{i}$ is replaced with the gradient of its expectation in \eqref{eq:est}. As depicted in Fig. \ref{fig:fig2}, the backward pass can be simply implemented with the hyperbolic tangent activation at the output layer.

\subsection{Centralized Training and Distributed Implementation}
The joint training strategy of optimizer and quantizer units is investigated here. The stochastic binarization layer properly handles the constraint in \eqref{eq:P21C1} of (P2.1) without loss of the optimality. Since the time sharing condition in Proposition \ref{lem:lem1} still holds for (P2.1), the primal-dual method in Section \ref{sec:sec3} is applied to train $2N$ DNNs for (P2.1). At the $t$-th iteration of the SGD training algorithm, the mini-batch updates for the DNN parameter $\vartheta^{[t]}\in\{\theta_{D,i}^{[t]},\theta_{Q,i}^{[t]}:\forall i\}$ and the dual variable $\lambda_{k}^{[t]}$ associated with \eqref{eq:P21C0} can be respectively written by
\begin{align}
    \vartheta^{[t]}&=\vartheta^{[t-1]}-\eta\bigg(\frac{1}{S}\sum_{\mathbf{a}\in\mathcal{S}}\nabla_{\vartheta}f(\mathbf{a},\phi_{D}(\tilde{\mathbf{a}}^{[t-1]};\theta_{D}^{[t-1]}))\nonumber\\
        &+\sum_{k=1}^{K}\mu_{k}^{[t-1]}\Big(\frac{1}{S}\sum_{\mathbf{a}\in\mathcal{S}}\nabla_{\vartheta}g_{k}(\mathbf{a},\phi_{D}(\tilde{\mathbf{a}}^{[t-1]};\theta_{D}^{[t-1]}))-G_{k}\Big)\bigg),\label{eq:p_up3}\\
    \lambda_{k}^{[t]}&=\bigg({\lambda_{k}}^{[t-1]}\nonumber\\
    &+\eta\Big(\frac{1}{S}\sum_{\mathbf{a}\in\mathcal{S}}g_{k}(\mathbf{a},\phi_{D}(\tilde{\mathbf{a}}^{[t-1]};\theta_{D}^{[t-1]}))-G_{k}\Big)\bigg)_{+},\label{eq:d_up3}
\end{align}
where
\begin{align}
    &\nabla_{\theta_{D,i}}\psi(\mathbf{a},\phi_{D}(\tilde{\mathbf{a}};\theta_{D}))\nonumber\\
    &=\nabla_{\phi_{D}(\tilde{\mathbf{a}};\theta_{D})}\psi(\mathbf{a},\phi_{D}(\tilde{\mathbf{a}};\theta_{D}))
        \cdot\nabla_{\theta_{D,i}}\phi_{D}(\tilde{\mathbf{a}};\theta_{D}),\nonumber\\
    &\nabla_{\theta_{Q,i}}\psi(\mathbf{a},\phi_{D}(\tilde{\mathbf{a}};\theta_{D}))\nonumber\\
    &=\nabla_{\phi_{Q,i}(\mathbf{a}_{i};\theta_{Q,i})}\psi(\mathbf{a},\phi_{D}(\tilde{\mathbf{a}};\theta_{D}))
        \cdot\nabla_{\tilde{\mathbf{a}}}\phi_{D}(\tilde{\mathbf{a}};\theta_{D})\cdot\nabla_{\theta_{Q,i}}\tilde{\mathbf{a}}\nonumber
\end{align}
for a function $\psi(\cdot)\in\{f(\cdot),g_{k}(\cdot):\forall k\}$. By \eqref{eq:p_up3} and \eqref{eq:d_up3}, Algorithm 1 can be also used in the joint training task of (P2.1). Note that the training of (P2.1) is an offline task performed in a centralized manner. The real-time computations at node $i$ are carried out by the individual DNNs $\phi_{D,i}(\mathbf{c}_{i};\theta_{D,i})$ and $\phi_{Q,i}(\mathbf{a}_{i};\theta_{Q,i})$ by means of the trained parameters stored in memory units. Therefore, node $i$ can calculate its solution $\mathbf{x}_{i}$ in a distributed manner, when operating. In practice, a backhaul link is realized by a reliable control channel, and the set of candidates for $B_{ij}$ is determined by service providers. In this setup, multiple DNNs are trained corresponding to candidates of $B_{ij}$ in advance. Then, this allows to each wireless node to select a suitable DNN from its memory unit for given backhaul capacity $B_{ij}$.

\begin{rem}
One can consider a distributed training strategy for (P2.1) where $2N$ DNNs deployed in individual wireless nodes are trained in a decentralized manner by exchanging relevant information for the training. To this end, individual DNNs forward their computation results with training samples, such as the cost function and the gradients of the weights and biases, both to the preceding and the proceeding layers. This results in a huge communication burden at each training iteration. Therefore, the proposed distributed deployment of jointly trained DNNs is practical for wireless nodes interconnected via capacity-limited backhaul links.
\end{rem}

\section{Applications to Wireless Resource Allocation}\label{sec:sec5}
In this section, numerical results are presented for demonstrating the efficacy of the proposed DL approaches to solve (P) in various wireless network applications. We consider two networking configurations: the C-MAC and the IFC systems. Such configurations are considered to be key enablers in next-generation communication networks such as LTE unlicensed and multi-cell systems. For each configuration, the power control is managed by the proposed DL strategies. In the C-MAC system, the optimality of the proposed DL methods, which is established analytically in Propositions 1 and 2, is verified numerically by investigating the optimal solution provided in \cite{RZhang:09} both for primal and dual domains. Next, the sum-rate maximization in the IFC, known to be non-convex, is addressed. The performance of the trained DNNs is tested and compared with a locally optimal solution obtained by the WMMSE algorithm \cite{QShi:11}. Finally, the proposed DL framework tackles the maximization of the minimum-rate performance in the IFC system. These results prove the viability of the DL technique in handling the optimization of non-smooth objective and constraints without additional reformulation.

\subsection{Application Scenarios}\label{sec:sec51}
\subsubsection{Cognitive multiple access channels}\label{sec:sec5B}
A multi-user uplink CR network \cite{RZhang:09} is considered. A group of $N$ secondary users (SUs), which share time and frequency resources with a licensed primary user (PU), desire to transmit their data simultaneously to a secondary base station (SBS). It is assumed that the PU, the SBS, and the SUs are all equipped with only a single antenna. Let $h_{i}$ and $g_{i}$ be the channel gain from SU $i$ $(i=1,\ldots,N)$ to the SBS and the PU, respectively. If the transmit power at SU $i$ is denoted by $p_{i}$, the average sum capacity maximization problem can be formulated as \cite{RZhang:09}
\begin{align}
    \text{(P3): }&\max_{\{p_{i}\}} \quad \mathbb{E}_{\mathbf{h},\mathbf{g}}\bigg[\log\bigg(1+\sum_{i=1}^{N}h_{i}p_{i}\bigg)\bigg]\nonumber\\
    &\text{subject to} \quad \mathbb{E}_{\mathbf{h},\mathbf{g}}[p_{i}]\leq P_{i},\ i=1,\ldots,N,\label{eq:P3C1}\\
    &\qquad \qquad \quad \mathbb{E}_{\mathbf{h},\mathbf{g}}\bigg[\sum_{i=1}^{N}g_{i}p_{i}\bigg]\leq \Gamma,\label{eq:P3C2}\\
    &\qquad \qquad \quad p_{i}\geq 0,\ i=1,\ldots,N,\label{eq:P3C3}
\end{align}
where $\mathbf{h}\triangleq\{h_{i}:\forall i\}$, $\mathbf{g}\triangleq\{g_{i}:\forall i\}$, $P_{i}$ and $\Gamma$ stand for the long-term transmit power budget at SU $i$ and the IT constraint for the PU, respectively. 
SU $i$ is responsible for 
computing the optimal transmit power $p_{i}^{\star}$. The optimal solution for (P3) is investigated in \cite{RZhang:09} using the Lagrange duality formulation and the ellipsoid algorithm. A centralized computation is necessary for the optimal solution in \cite{RZhang:09} under the assumption that the global channel gains $\mathbf{h}$ and $\mathbf{g}$ are perfectly known to SU $i$. 

Note that (P3) is regarded as a special case of (P) by setting $\mathbf{a}=\{\mathbf{h},\mathbf{g}\}\in\mathbb{R}^{2N}$ and $\mathbf{x}=\{x_{i}:\forall i\}\in\mathbb{R}^{N}$ with $x_{i}=p_{i}$. In the centralized approach, SU $i$ forwards the local information $\mathbf{a}_{i}=\{h_{i},g_{i}\}\in\mathbb{R}^{2}$ to neighboring SU $j~(j\neq i)$. The deterministic constraint in \eqref{eq:P3C3} can be simply handled by the ReLU activation at the output layer, which yields a projection onto a non-negative feasible space with $p_{i}\geq 0$. The overall network solution is obtained using $\mathbf{x}=\phi_{C}(\mathbf{a};\theta_{C})$. For the distributed implementation, SU $i$ first obtains the quantization $\mathbf{v}_{i}=\phi_{Q,i}(\mathbf{a}_{i};\theta_{Q,i})$ for transfer to other SUs and, in turn, calculates its solution using $x_{i}=\phi_{D,i}(\mathbf{c}_{i};\theta_{D,i})$.

\subsubsection{Interference channels}\label{sec:sec5C}
In an IFC scenario, there are $N$ transmitters which send their respective messages to the corresponding receivers at the same time over the same frequency band. The channel state information between a pair of transmitter $i$ and receiver $j$ is denoted by $h_{ij}$. We consider two different power control challenges: the sum rate maximization and the minimum rate maximization. First, the sum rate maximization is formulated as
\begin{align}
    \text{(P4): }&\max_{\{p_{i}\}}\quad \mathbb{E}_{\mathbf{h}}\bigg[\sum_{i=1}^{N}\log\bigg(1+\frac{h_{ii}p_{i}}{1+\sum_{j\neq i}h_{ji}p_{j}}\bigg)\bigg]\nonumber\\
    &\text{subject to} \quad \mathbb{E}_{\mathbf{h}}[p_{i}]\leq P_{A,i},\ i=1,\ldots,N,\label{eq:P4C1}\\
    &\qquad \qquad \quad 0\leq p_{i}\leq P_{P,i},\ i=1,\ldots,N.\label{eq:P4C2}
\end{align}
With a slight abuse of notations, $\mathbf{h}\triangleq\{h_{ij}:\forall (i,j)\}$ and $p_{i}$ can be defined as the collection of the channel gains and the transmit power at transmitter $i$, respectively. Two different transmit power constraints are taken into account in (P4): the average power constraint $P_{A,i}$ in \eqref{eq:P4C1} and the peak power constraint $P_{P,i}$ in \eqref{eq:P4C2}. It is known in \cite{HSun:18} that (P4) is non-convex and NP hard.

For a special case of $P_{A,i}=P_{P,i}$, the average power constraint in \eqref{eq:P4C1} is ignored. In such a case, a local optimal solution for (P4) is obtained by the WMMSE algorithm \cite{QShi:11}. The DL techniques have been recently applied to solve (P4) for the special case $P_{A,i}=P_{P,i}$ \cite{HSun:18,WLee:18a}. Those methods, however, depend on the global CSI $\mathbf{h}$, thereby involving centralized computations for obtaining the power control solution. A naive distributed scheme has been proposed in \cite{WLee:18a} where zero input vector is applied to the trained DNN for the unknown information, while its performance is not sufficient in practice. Furthermore, in the case of $P_{A,i}\leq P_{P,i}$, there is no generic algorithm for (P4) even with centralized computations.

Next, the minimum rate maximization is expressed as
\begin{align}
    \text{(P5): }&\max_{\{p_{i}\}}\quad \mathbb{E}_{\mathbf{h}}\bigg[\min_{i}\log\bigg(1+\frac{h_{ii}p_{i}}{1+\sum_{j\neq i}h_{ji}p_{j}}\bigg)\bigg]\nonumber\\
    &\text{subject to} \quad \mathbb{E}_{\mathbf{h}}[p_{i}]\leq P_{A,i},\ i=1,\ldots,N,\nonumber\\
    &\qquad \qquad \quad 0\leq p_{i}\leq P_{P,i},\ i=1,\ldots,N.\nonumber
\end{align}
This, in general, turns out to be a non-convex and non-smooth optimization problem. If $P_{A,i}=P_{P,i}$, the globally optimal solution can be obtained from an iterative algorithm in \cite{DCai:12}. However, it requires to share the power control solution of all other cell transmitters until convergence, resulting in the use of infinite backhaul links. Furthermore, in a general case of $P_{A,i}\leq P_{P,i}$, no efficient solution for (P5) is available. In optimization theory, additional transformation, such as epigraph form \cite{Boyd:04} and Perron-Frobenius theory \cite{DCai:12}, are necessary to reformulate an analytically intractable objective function of (P5). Rigorous analytical processing of equivalent reformulations for such a purpose typically result in additional auxiliary optimization variables, i.e., the increase in the dimension of a solution space. Proper address of (P5) with the proposed DL approaches shows the power of the DNNs for tackling non-smooth optimization problems without traditional reformulation techniques.

The proposed DL framework is applied to solve (P4) and (P5). Transmitter $i$ now becomes a computing node that yields its power control strategy $x_{i}=p_{i}$ from the local CSI $\mathbf{h}_{i}\triangleq\{h_{ji}:\forall j\}$.\footnote{The channel gain $h_{ji}$ can be estimated at receiver $i$ based on standard channel estimation processes. Then, receiver $i$ can inform $\mathbf{h}_{i}$ to its corresponding transmitter through reliable uplink feedback channels \cite{WLee:18a}.} Similar to the C-MAC problem (P3), both the centralized and the decentralized DL approaches are readily extended to (P4) and (P5).

\subsection{Numerical Results}
We test the proposed DL framework for solving the power control problems (P3)-(P5). The details of the implementation and simulation setup are described first. 
In the centralized approach, the DNN is constructed with 4 hidden layers, each of the output dimension $10N$. In the distributed approach, the optimizer DNN consists of 3 hidden layers, each of the output dimension $10N$, while the quantizer DNN is a single hidden layer neural network of the dimension $10N$. The ReLU activation $\text{ReLU}(x)\triangleq(x)_{+}$ is employed for all hidden layers. Since, in both approaches, each node is equipped with 4 hidden layers of the same ReLU activation, they are implemented with comparable computation complexities at each node. For efficient training, the batch normalization technique is applied at each layer~\cite{Ioffe:15}.\footnote{Although a more sophisticated structure of the DNN for the applications given in Section \ref{sec:sec51} can be constructed, the optimization of the DNN structure is out of scope.} In the proposed training strategy, the updates of the DNN parameters and the dual variables are carried out by the state-of-the-art SGD algorithm based on Adam optimizer \cite{Kingma:15}.
The learning rate and the mini-batch size are $\eta=5\times10^{-5}$ and $S=5\times10^{3}$, respectively. At each iteration, mini-batch samples, i.e., the channel gains, are independently generated from the exponential distribution with unit mean.\footnote{The proposed DL approaches are examined to work well in practical distance-based channel models, and similar trends are observed as in a simple Rayleigh fading setup.} The number of iterations for the Adam optimizer is set to $5\times 10^5$, and thus the total of $2.5\times 10^9$ samples are applied during the training. The DNN parameters are initialized according to the Xavier initialization \cite{XGlorot:10}. The initial weight matrices are randomly generated from zero-mean Gaussian random variable matrices with the variance normalized by the input dimension, while all elements of the initial bias vectors are fixed to $0.01$. The initial value of each dual variable is $\lambda_{k}^{[0]}=0$. During the training, the performance of the DNNs is assessed over the validation set of size $10^{6}$ samples randomly generated from training mini-batch sets. Finally, the performance of the trained DNN is examined over the testing set with $10^{4}$ samples which are not known to the DNNs during the training and the validation. The proposed training algorithms are implemented in Python 3.6 with TensorFlow 1.4.0.

\subsubsection{Cognitive multiple access channels}
\begin{figure}
\begin{center}
\includegraphics[width=\linewidth]{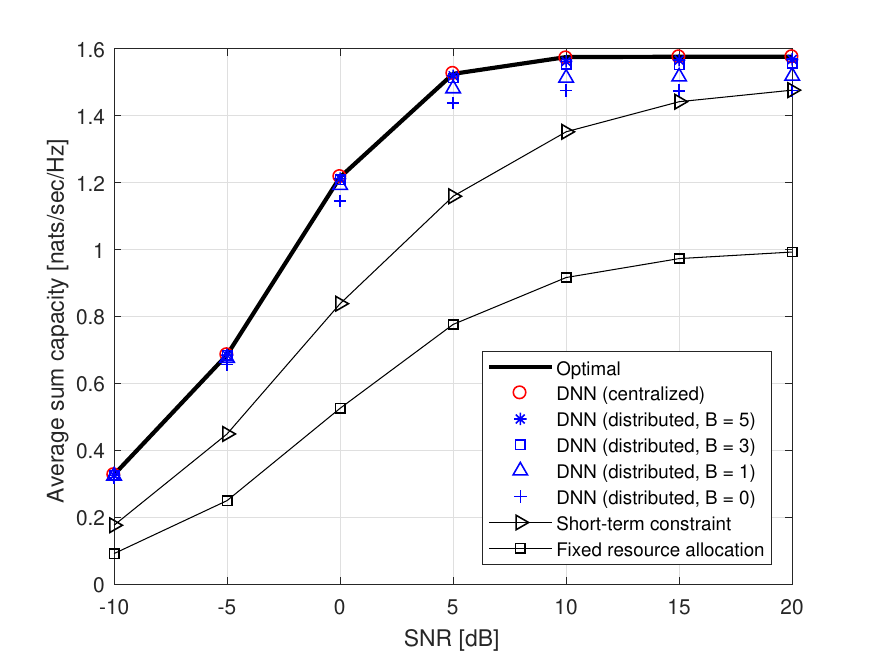}
\end{center}
\caption{Average sum capacity performance with respect to SNR.}
\label{fig:fig3}
\end{figure}

The performance of the proposed DL framework for (P3) is presented with $N=2$, $P_{i}=P$, $\forall i$, $\Gamma=1$, and $B_{ij}=B$, $\forall (i,j)$. 
Fig. \ref{fig:fig3} shows the average sum capacity performance of the proposed DL framework for (P3) evaluated over the testing set. The signal-to-noise ratio (SNR) is defined as $\text{SNR}=P$. The following three baseline schemes provided in \cite{RZhang:09} are considered for comparison:
\begin{itemize}
\item \textit{Optimal:} The optimal solution for (P3) is computed from the algorithm in \cite{RZhang:09}.
\item \textit{Short-term constraint:} (P3) is optimally solved with the short-term constraints $p_{i}\leq P$ and $\sum_{i=1}^{N}g_{i}p_{i}\leq \Gamma$, instead of \eqref{eq:P3C1} and \eqref{eq:P3C2}.
\item \textit{Fixed resource allocation:} A simple power control strategy is employed as $p_{i}=\max\{P,\frac{\Gamma}{g_{i}}\}$.
\end{itemize}
It is observed that the proposed centralized DNN achieves the optimal average sum-capacity performance. 
The proposed distributed DL method shows a slight performance gap from the optimal scheme, albeit shrinking as the backhaul capacity $B$ increases. In this example, $B=3$ is shown to reach almost identical performance of the optimal scheme. The proposed distributed DL framework can reduce the network delay and the backhaul signaling overhead for sharing the perfect global CSI in existing optimization algorithm. Hence, the DL technique proves powerful in addressing the network optimization tasks in the fast fading environment. The distributed DNN with $B=0$, where the coordination among nodes is not available, presents a performance close to the optimal solution. During the training step, each of distributed DNNs observes average behavior of other DNNs via the stochastic update rules in (\ref{eq:p_up3}) and (\ref{eq:d_up3}). As a result, although the coordination is not allowed, the distributed DL technique can learn statistical nature of the overall wireless systems and outperforms other baseline schemes.

\begin{figure}
\centering
    \subfigure[Primal convergence]{
        \includegraphics[width=\linewidth]{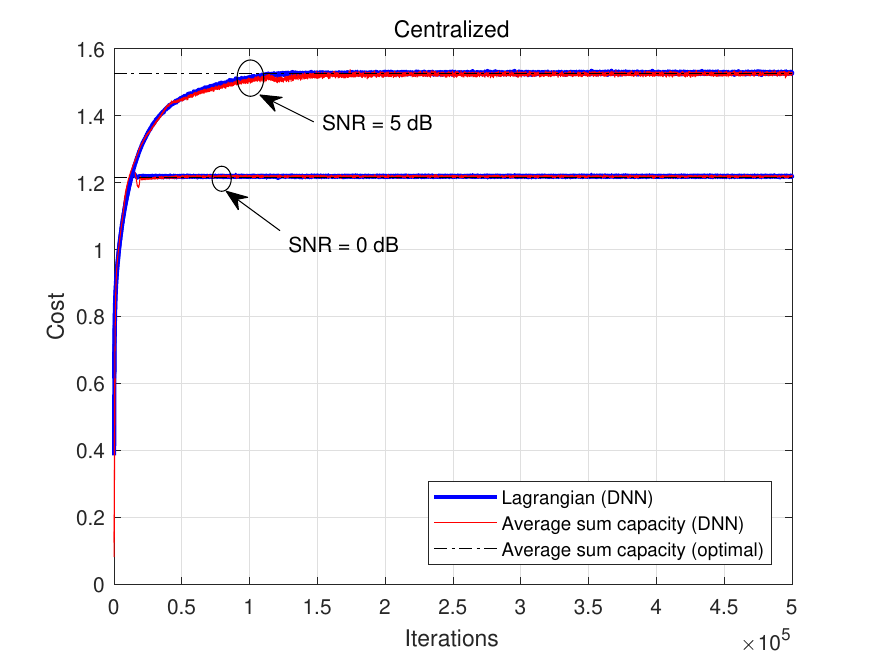}\label{fig:fig4a}
    }
    \subfigure[Dual convergence]{
        \includegraphics[width=\linewidth]{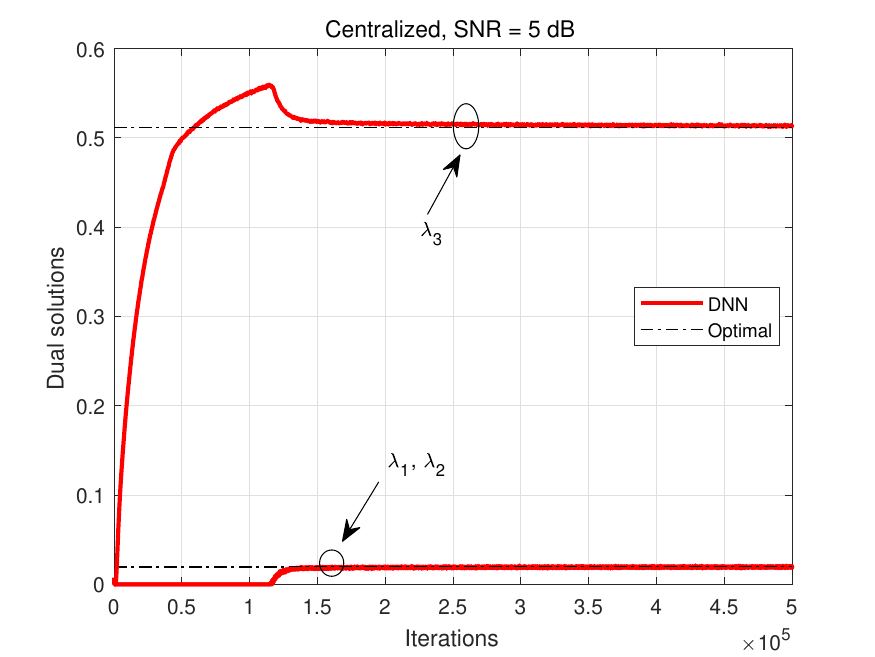}\label{fig:fig4b}
    }
    \subfigure[Primal feasibility]{
        \includegraphics[width=\linewidth]{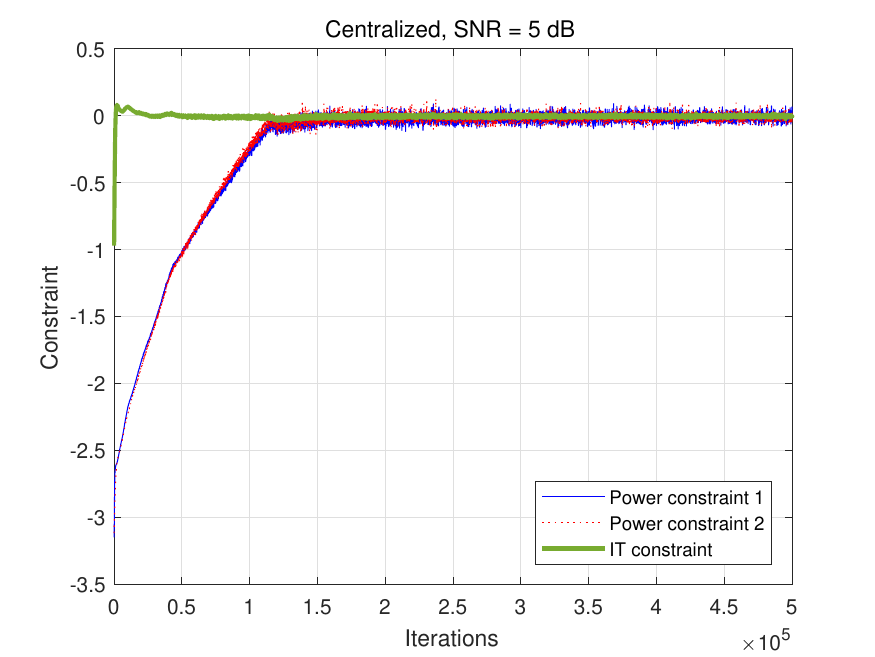}\label{fig:fig4c}
    }
    \caption{Convergence behavior of the proposed centralized approach.}
    \label{fig:fig4}
\end{figure}

Fig.~\ref{fig:fig4} illustrates the convergence behaviors of the proposed training algorithm for the centralized approach. Fig.~\ref{fig:fig4a} depicts the convergence behavior of the primal variable, i.e., the DNN parameter $\theta_{C}$, in the centralized approach for $\text{SNR}=0$ and $5\ \text{dB}$. The Lagrangian and the average sum capacity of the DNN converge to the optimal performance consistently over all SNR regimes, indicating that the strong duality addressed in Proposition 2 indeed holds. Fig.~\ref{fig:fig4b} shows the convergence of the dual variable obtained from the proposed training algorithm for $\text{SNR}=5\ \text{dB}$. Two dual variables $\lambda_{1}$ and $\lambda_{2}$ are associated with the long-term power constraint in \eqref{eq:P3C1} and $\lambda_{3}$ corresponds to the IT constraint in \eqref{eq:P3C2}. The proposed training algorithm can find the optimal dual variable. The dual variables $\lambda_{1}$ and $\lambda_{2}$ do not update before fast increases around the $1.1\times10^5$-th training iteration. By contrast, $\lambda_{3}$ grows rapidly from the beginning of the training iteration and then tunes to the optimal solution. This can be explained from Fig. \ref{fig:fig4c} which examines the primal feasibility by presenting the power constraints $\mathbb{E}_{\mathbf{h},\mathbf{g}}[p_{i}]-P$ for $i=1,2$ and the IT constraint $\mathbb{E}_{\mathbf{h},\mathbf{g}}[\sum_{i=1}^{N}g_{i}p_{i}]-\Gamma$. The power constraints initially remain inactive as $\mathbb{E}_{\mathbf{h},\mathbf{g}}[p_{i}]-P_{A}<0$ for the complementary slackness $\lambda_{i}(\mathbb{E}_{\mathbf{h},\mathbf{g}}[p_{i}]-P_{A})=0$ leading to null corresponding dual variables \cite{Boyd:04}. If the power constraints now reach their bound $P$, the dual variables $\lambda_{1}$ and $\lambda_{2}$ become positive as shown in Fig. \ref{fig:fig4b} to control the primal feasibility. Meanwhile, the DNN violates the IT constraint at the beginning of the training iterations as illustrated in Fig. \ref{fig:fig4c}. Therefore, the dual variable $\lambda_{3}$ quickly increases and becomes larger than its optimal value to regularize the IT constraint. As the DNN is trained and gets feasible, $\lambda_{3}$ decreases and converges to the optimal solution. As shown in Figs. \ref{fig:fig4b} and \ref{fig:fig4c} the DNN satisfies the complementary slackness, which ensures the zero duality gap \cite{Boyd:04}.

\begin{figure}
\centering
    \subfigure[Primal convergence]{
        \includegraphics[width=\linewidth]{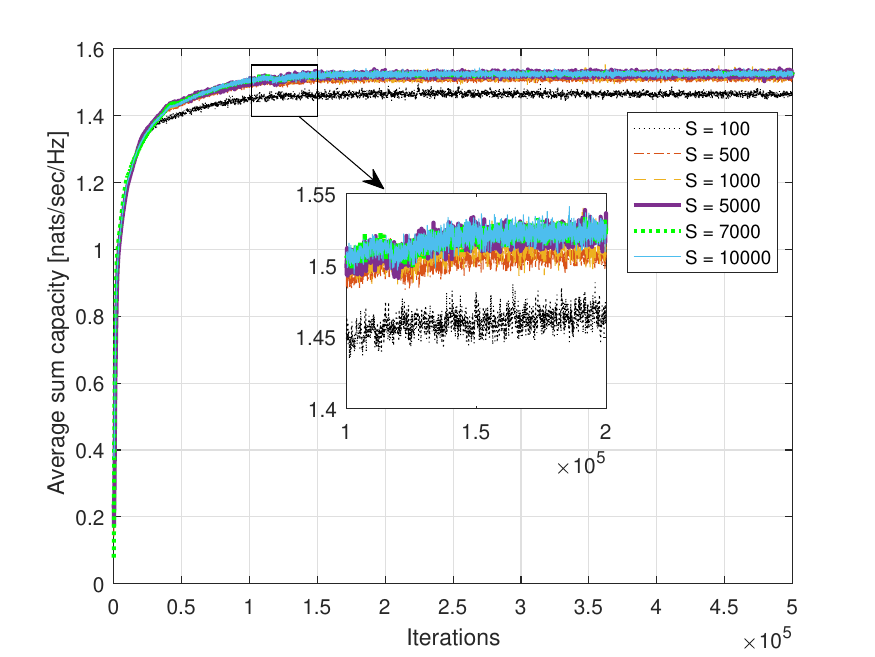}\label{fig_R1_C6a}
    }
    \subfigure[Dual convergence]{
        \includegraphics[width=\linewidth]{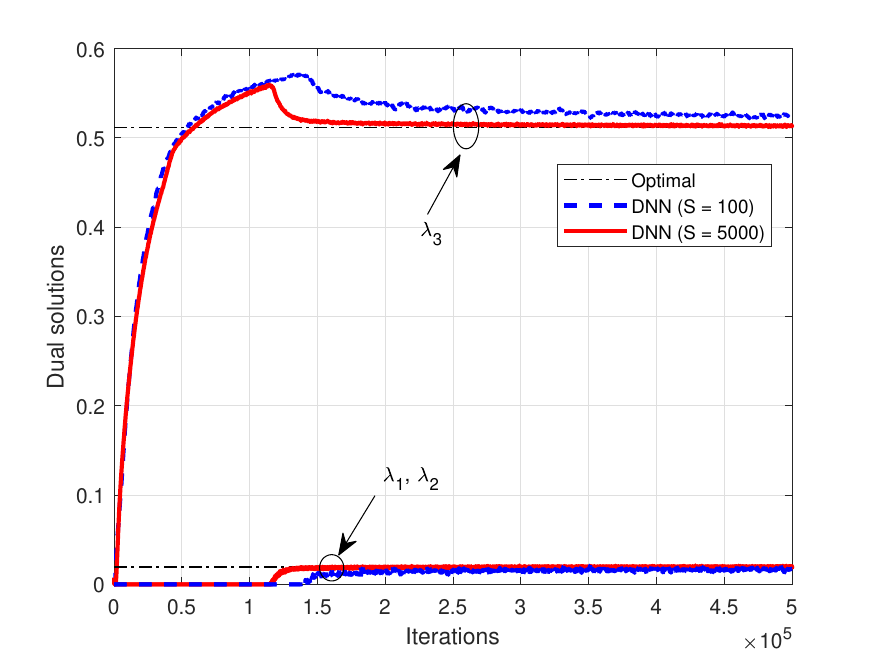}\label{fig_R1_C6b}
    }
    \caption{Convergence behavior of the proposed centralized approach with different mini-batch sizes $S$.}
    \label{fig:fig_R1_C6}
\end{figure}

Fig. \ref{fig:fig_R1_C6} shows the convergence behavior of the proposed centralized DL approach at $\text{SNR}=5\ \text{dB}$ for various mini-batch sizes $S$. Fig. \ref{fig_R1_C6a} shows the primal objective convergence by plotting the average sum capacity with respect to training iterations. The DNN trained with small batch sizes, e.g., $S=100, 500$, and $1000$, exhibits the performance degradation. Small mini-batch sizes cause inaccurate approximation of the average cost and constraints functions in (\ref{eq:p_up2}) and (\ref{eq:d_up2}), resulting in incorrect estimates on dual variables as illustrated in Fig. \ref{fig_R1_C6b} presenting the dual objective convergence. On the other hand, a large $S$ incurs expensive computations at each training iteration. Considering convergence behavior and computational complexity, $S = 5000$ is a suitable choice for simulation.

\begin{figure}
\centering
    \subfigure[Primal convergence]{
        \includegraphics[width=\linewidth]{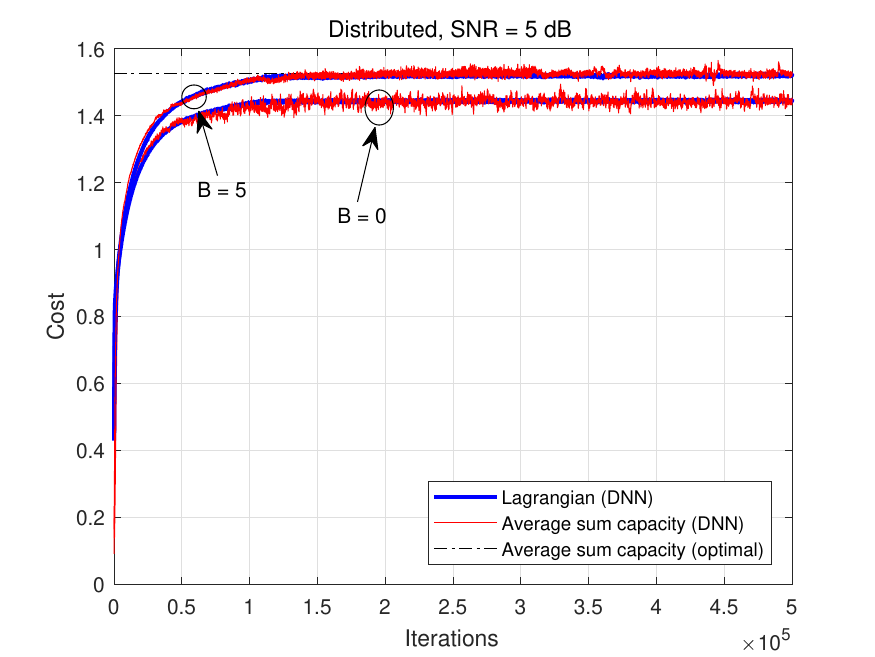}\label{fig:fig5a}
    }
    \subfigure[Dual convergence]{
        \includegraphics[width=\linewidth]{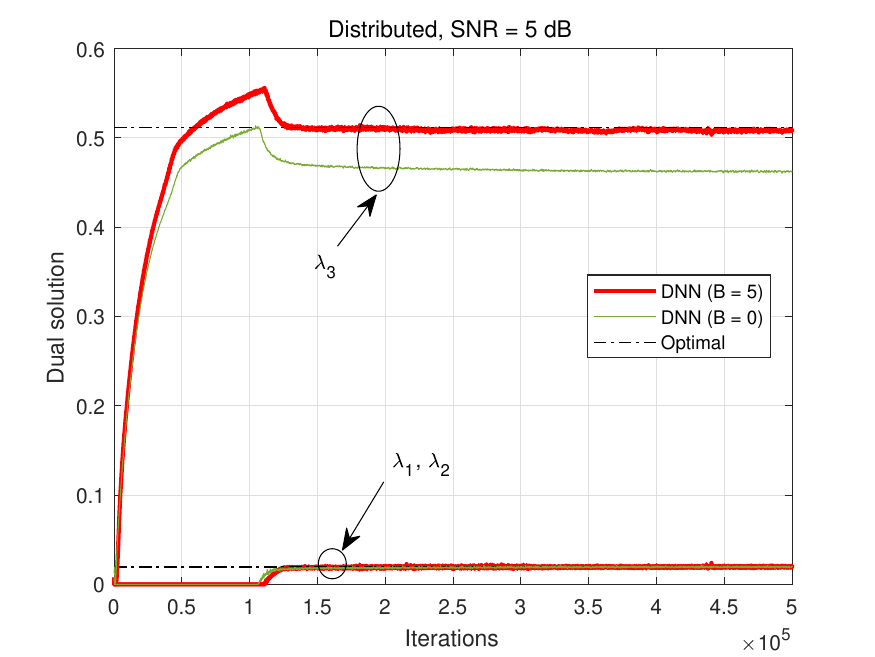}\label{fig:fig5b}
    }
    \caption{Convergence behavior of the proposed distributed approach.}
    \label{fig:fig5}
\end{figure}

Fig.~\ref{fig:fig5} exhibits the convergence of the distributed approach at $\text{SNR}=5\ \text{dB}$ for $B=0$ and $5$ bits. For $B=5$, the proposed distributed DL method converges to the optimal performance of \cite{RZhang:09}, with the infinite-capacity backhaul, in the primal and dual domains. However, if the information sharing among SUs is not allowed, i.e., $B=0$, it fails to identify the optimal dual variable $\lambda_{3}$ for the IT constraint as illustrated in Fig. \ref{fig:fig5b}. This causes a performance loss compared to the optimal scheme and indicates that the quantization and the data sharing are crucial features in the distributed implementation.

\begin{figure}
\begin{center}
\includegraphics[width=\linewidth]{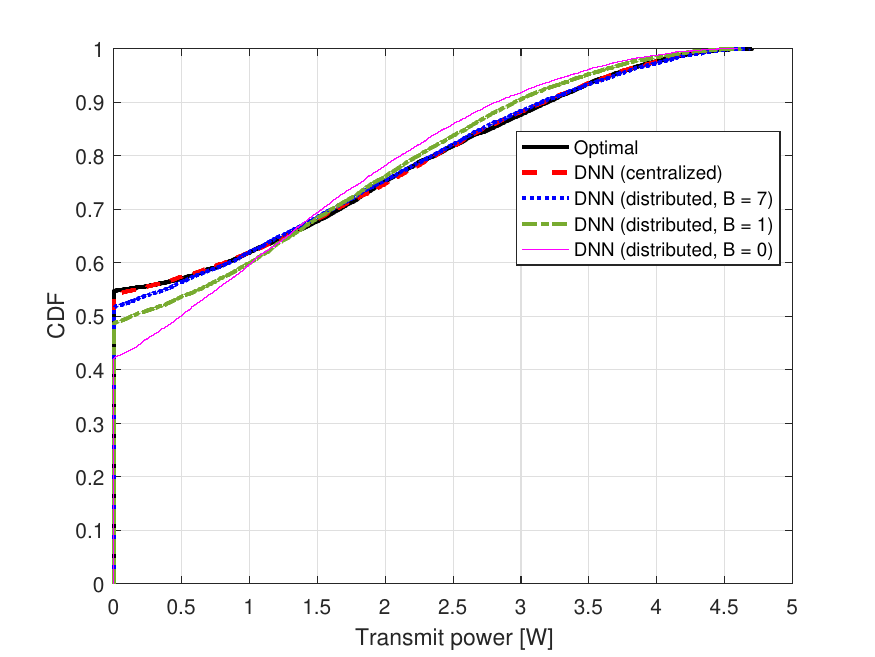}
\end{center}
\caption{CDF of the transmit power at the SUs for different backhaul link capacity.}
\label{fig:fig_R1_C8}
\end{figure}

To see the impact of the backhaul link capacity $B$ on the performance of the DNN, Fig. \ref{fig:fig_R1_C8} plots the cumulative density function (CDF) of the transmit power at the SUs at $\text{SNR}=0\ \text{dB}$. The centralized DNN yields the identical CDF to the optimum \cite{RZhang:09}. The CDF of the transmit power computed by the distributed DNN approaches to the optimal solution as $B$ increases. When $B=0$, the distributed DNN is likely to yield higher transmit power than the optimal algorithm, which undermines the sum-rate performance. This results from egoistic solutions of separate DNNs deployed in individual SUs for the lack of information shared among them. Therefore, SUs transmit their data signals with the maximum allowable power while meeting constraints (\ref{eq:P3C1}) and (\ref{eq:P3C2}).

\begin{figure}
\centering
    \subfigure[Number of hidden layers with the dimension $10N$]{
        \includegraphics[width=\linewidth]{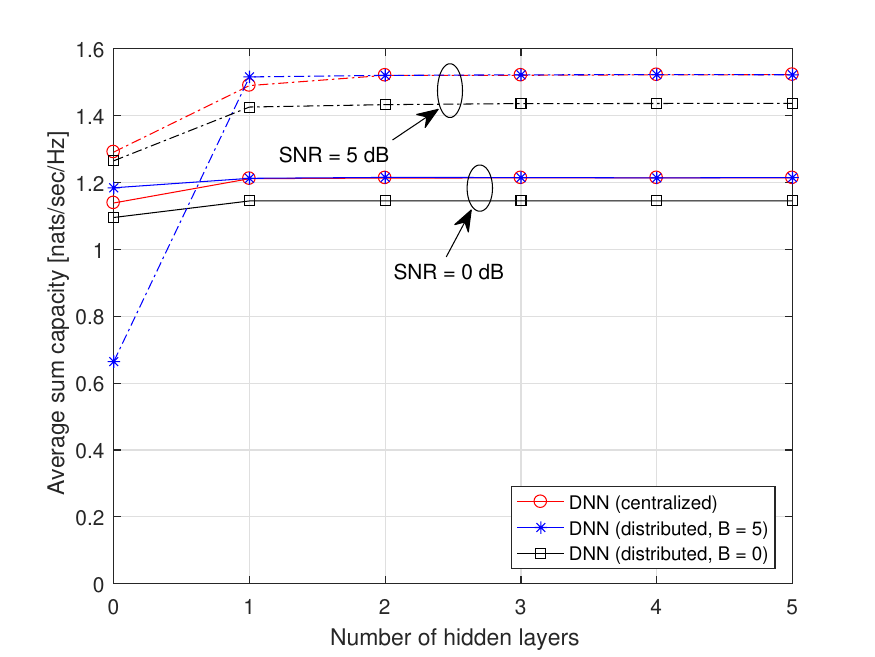}\label{fig:fig_R1_C7a}
    }
    \subfigure[Dimension of hidden layers with four hidden layers]{
        \includegraphics[width=\linewidth]{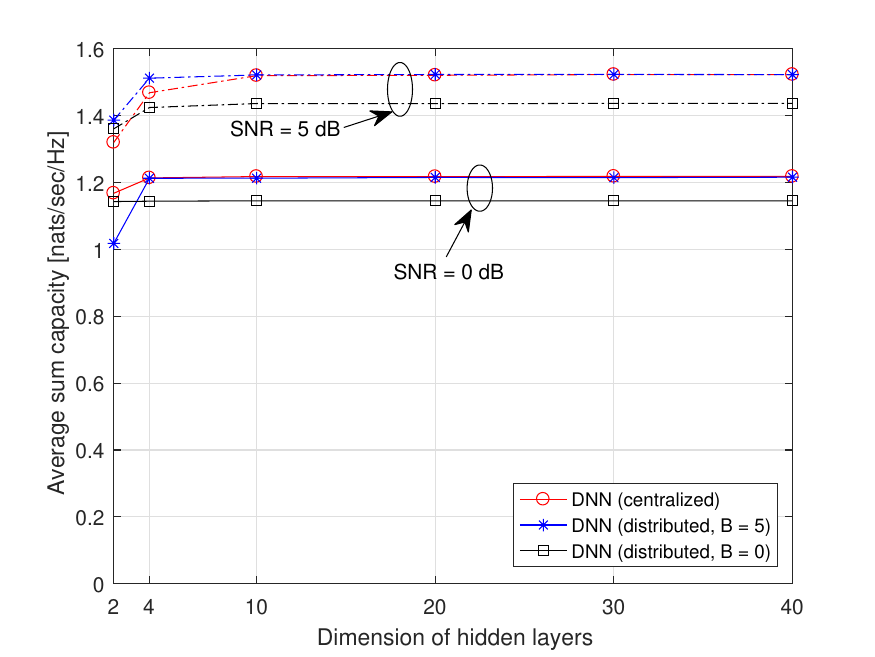}\label{fig:fig_R1_C7b}
    }
    \caption{Impact of the DNN structure in the proposed centralized approach.}
    \label{fig:fig_R1_C7}
\end{figure}

Fig. \ref{fig:fig_R1_C7} presents the average sum capacity performance of the proposed DNN in the C-MAC setup (P3) for variations of the DNN structure, i.e., the number and the dimension of hidden layers. Fig. \ref{fig:fig_R1_C7a} plots the average sum capacity performance as a function of the number of hidden layers with the dimension $10N=20$. The number of hidden layers equal to zero corresponds to the DNN consisting of only input and output layers. It shows that the average sum capacity performance first increases as the number of hidden layers increases and saturates at four hidden layers. A similar observation is obtained from Fig. \ref{fig:fig_R1_C7b} for different hidden layer dimensions with four hidden layers. Based on empirical results, we have constructed a DNN with four hidden layers, each with dimension $20$.

\subsubsection{Interference channels}

\begin{figure}
\centering
    \subfigure[Average sum rate with respect to SNR]{
        \includegraphics[width=\linewidth]{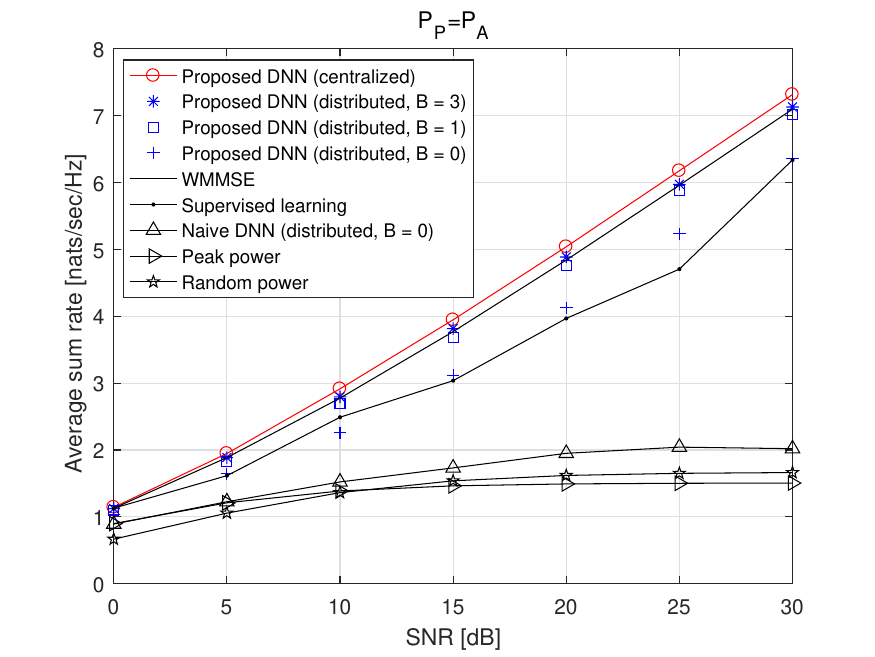}\label{fig:fig6a}
    }
    \subfigure[Average sum rate with respect to $B$]{
        \includegraphics[width=\linewidth]{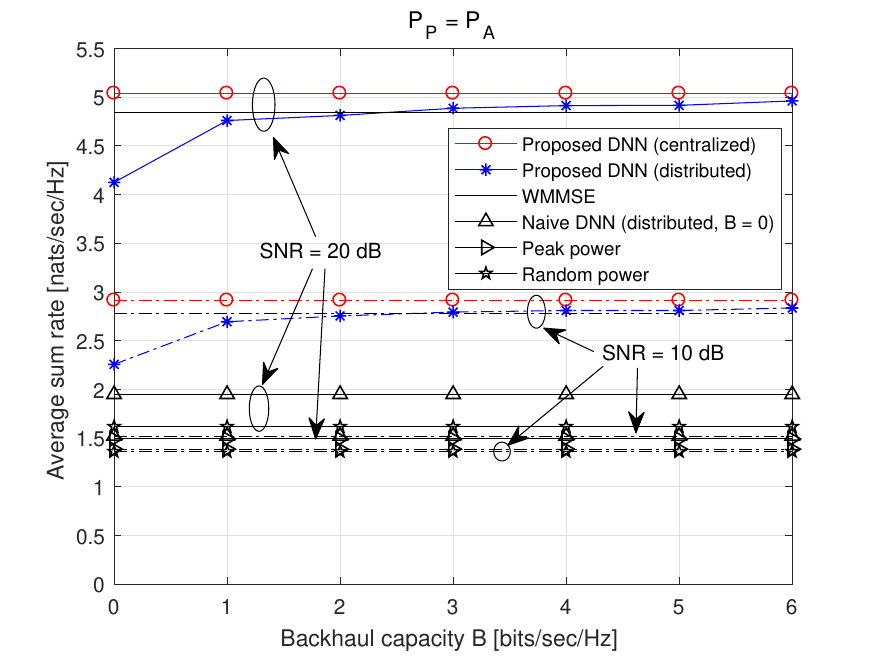}\label{fig:fig6b}
    }
    \caption{Average sum rate performance of the proposed DL framework with $P_{A}=P_{P}$.}
    \label{fig:fig6}
\end{figure}

For the IFC, we set $N=3$, $P_{A,i}=P_{A}$, $\forall i$, $P_{P,i}=P_{P}$, $\forall i$, and $B_{ij}=B$, $\forall (i,j)$. First, we focus on the sum rate maximization problem (P4). Fig.~\ref{fig:fig6} shows the average sum rate of the proposed DL methods for the case of $P_{A}=P_{P}$ where the average power constraint in \eqref{eq:P4C1} becomes inactive. The SNR is set to $\text{SNR}=P_{A}$. The following baseline schemes are considered as references, while those are only applicable to the case of $P_{A}=P_{P}$.
\begin{itemize}
\item \textit{WMMSE:} A locally optimal solution for (P4) is obtained by using the WMMSE power control algorithm \cite{QShi:11} with the initial point $p_{i}=P_{A}$, $\forall i$.
\item \textit{Supervised learning:} The DNN is trained to use supervised learning with the solution of the WMMSE algorithm as a label for the training set \cite{HSun:18}.
\item \textit{Naive DNN:} No data sharing is allowed among the transmitters, i.e., $B=0$. Then, the DNN trained from the proposed centralized approach is directly implemented at each transmitter by inputting zeros for the unknown channel gains of other transmitters \cite{WLee:18a}.
\item \textit{Peak power:} The transmitters adapt the peak power transmission strategy $p_{i}=P_{P}$,~$\forall i$.
\item \textit{Random power:} The transmit power $p_{i}$ is distributed uniformly in $[0,P_{P}]$.
\end{itemize}
Fig. \ref{fig:fig6a} illustrates the average sum rate performance with respect to SNR. Note that the proposed centralized DL technique outperforms a locally optimal WMMSE algorithm for all SNR regimes. This evidences that the non-convexity of the optimization problem, which prevents from finding an efficient solution in a simple way, would be handled by the DL-based data-driven approach. By contrast, the supervised learning strategy in \cite{HSun:18} exhibits slightly degraded performance compared to the WMMSE algorithm.\footnote{It has been reported in \cite{HSun:18} that when the dimension of hidden layers is large as 200, the supervised learning strategy shows a small performance loss compared to the WMMSE algorithm at the low SNR regime. However, in this work, a simpler DNN structure is employed and trained at the high SNR.} Also, $B=1$ suffices for the proposed distributed approach to attain the performance of the WMMSE algorithm, which requires the sharing of the perfect CSI among the transmitters with the infinite-capacity backhaul. The proposed distributed DNN with $B=0$ and the naive distributed DNN baseline in \cite{WLee:18a} both do not allow the information sharing among the transmitters. Nevertheless, the proposed distributed algorithm with $B=0$ outperforms the naive distributed DNN in \cite{WLee:18a}. Unlike the technique used in \cite{WLee:18a} that straightforwardly utilizes the DNN trained with the perfect global CSI, the proposed approach trains multiple DNNs without the global CSI for the distributed implementation. Thus, the proposed distributed DL technique excels in learning the nature of the sum rate maximization only with the local CSI. Fig. \ref{fig:fig6b} plots the average sum rate of the network with respect to the backhaul capacity $B$ with $P_{A}=P_{P}$ for $\text{SNR}=10$ and $20\ \text{dB}$. The distributed DNN narrows the performance gap from the centralized method as the backhaul capacity $B$ increases consistently over all SNR regimes. As reported in \cite{HSun:18}, the proposed DL schemes can save the computational time of the WMMSE algorithm. The CPU running times of centralized and distributed DNNs are only $4\ \%$ and $6.7\ \%$ of the WMMSE algorithm, respectively. Such computational time saving results from real-time calculations of the DNNs evaluating simple matrix multiplications, whereas the WMMSE algorithm is subject to iterative calculations.

\begin{figure}
\centering
    \subfigure[Average sum rate with respect to SNR]{
        \includegraphics[width=\linewidth]{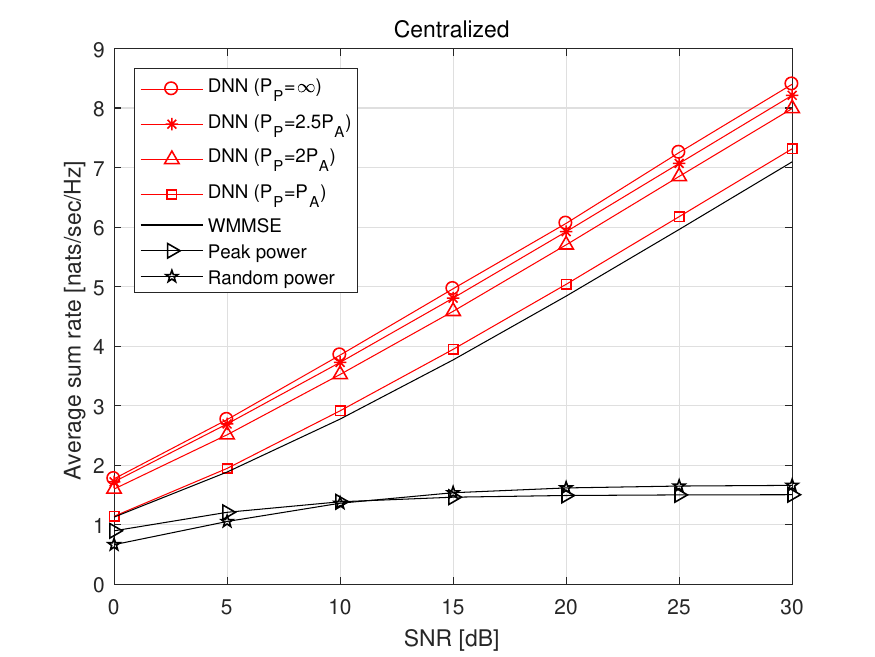}\label{fig:fig7a}
    }
    \subfigure[Average sum rate with respect to $B$]{
        \includegraphics[width=\linewidth]{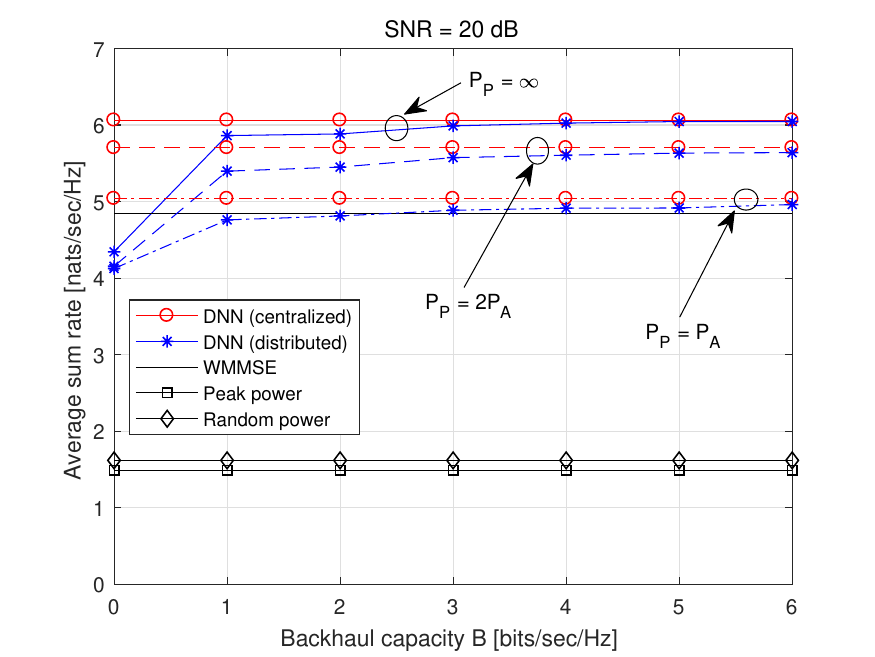}\label{fig:fig7b}
    }
    \caption{Average sum rate performance of the proposed DL framework with different $P_{P}$.}
    \label{fig:fig7}
\end{figure}
Fig. \ref{fig:fig7} presents numerical results for a general case that $P_{A}\leq P_{P}$, i.e., both the peak power and the average power constraints are considered together for (P4). Fig. \ref{fig:fig7a} depicts the average sum rate of the centralized approach with respect to SNR for different $P_{P}$. The average sum rate increases as the peak power grows. For all simulated configurations of $P_{A}$, the peak power constraint $P_{P}=2.5P_{A}$ achieves almost identical performance without the unbounded case, i.e., $P_{P}=\infty$. Fig. \ref{fig:fig7b} illustrates the average sum rate with respect to $B$ at $\text{SNR}=20\ \text{dB}$. The performance of the distributed DNN improves monotonically as $B$ increases, indicating that the proposed approach works well under both the average and the peak power constraints.

\begin{figure}
\centering
    \subfigure[Average max-min rate with respect to SNR]{
        \includegraphics[width=\linewidth]{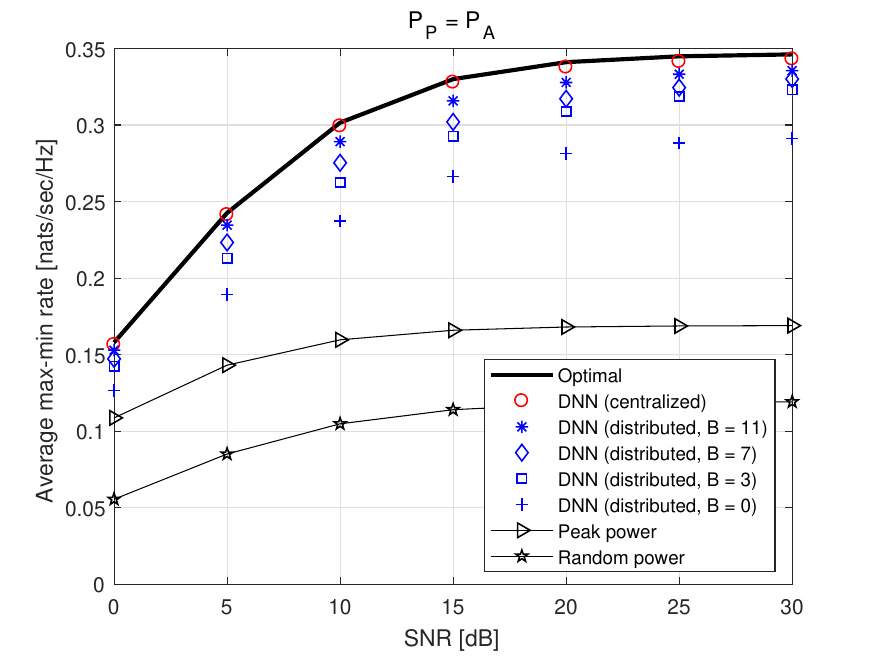}\label{fig:fig8a}
    }
    \subfigure[Average max-min rate with respect to $B$]{
        \includegraphics[width=\linewidth]{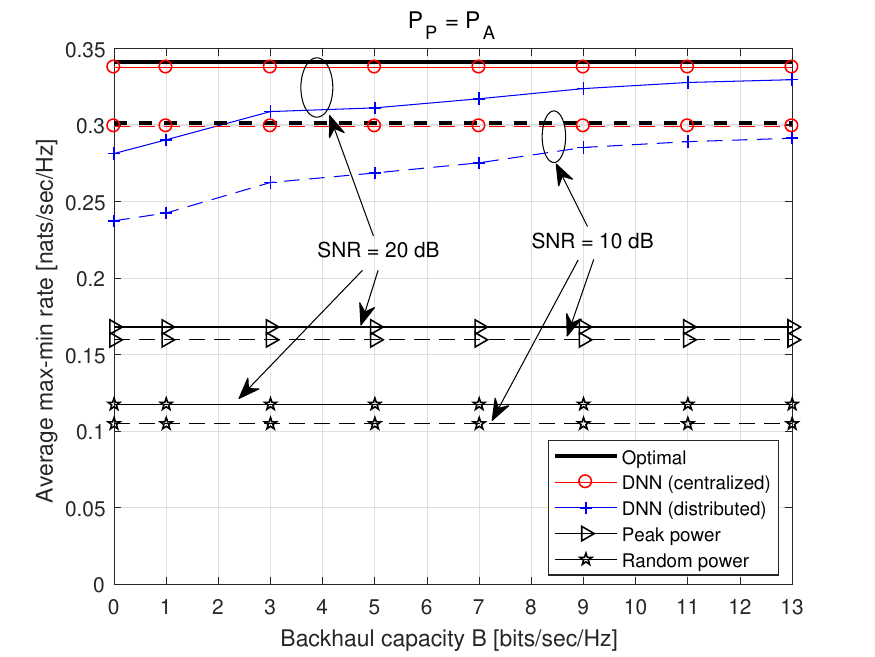}\label{fig:fig8b}
    }
    \caption{Average max-min rate performance of the proposed DL framework with $P_{A}=P_{P}$.}
    \label{fig:fig8}
\end{figure}

Next, the minimum rate maximization in (P5) is investigated. In this configuration, a more complicated DNN structure is employed with the increased number of hidden layers and the output dimension than the sum rate maximization case. For the centralized approach, the DNN is realized by 5 hidden layers with the output dimension $20N$. In addition, the optimizer DNN of the distributed case uses 4 hidden layers with the output dimension $20N$, while the quantizer DNN is constructed by a single hidden layer neural network of the dimension $20N$.

Fig. \ref{fig:fig8} depicts the average maximized minimum (max-min) rate for the case of $P_{A}=P_{P}$, i.e., the average power constraint is ignored. The optimal performance could be obtained from the method in \cite{DCai:12}. In Fig. \ref{fig:fig8a}, the average max-min rate is presented with respect to SNR with $P_{A}=P_{P}$. For all SNR regimes, the centralized DL presents a near-optimal performance. Even in the case of no cooperation ($B=0$), the proposed distributed DL outperforms the baseline power allocation schemes significantly. Compared to the sum rate maximization, the distributed approach in the max-min power control application in (P5) requires a higher backhaul capacity to match the performance of the centralized DL method, in particular, in moderate to high SNR regimes. Thus, the performance of the DNN at SNR = 10 and 20 dB is a primary focus, and Fig. \ref{fig:fig8b} illustrates the average max-min rate with respect to the backhaul link capacity with $P_{A}=P_{P}$ for $\text{SNR}=10$ and $20\ \text{dB}$. As shown in the sum rate maximization, the distributed approach gradually reduces the gap from the centralized method as $B$ grows.

\begin{figure}
\centering
    \subfigure[Average max-min rate with respect to SNR]{
        \includegraphics[width=\linewidth]{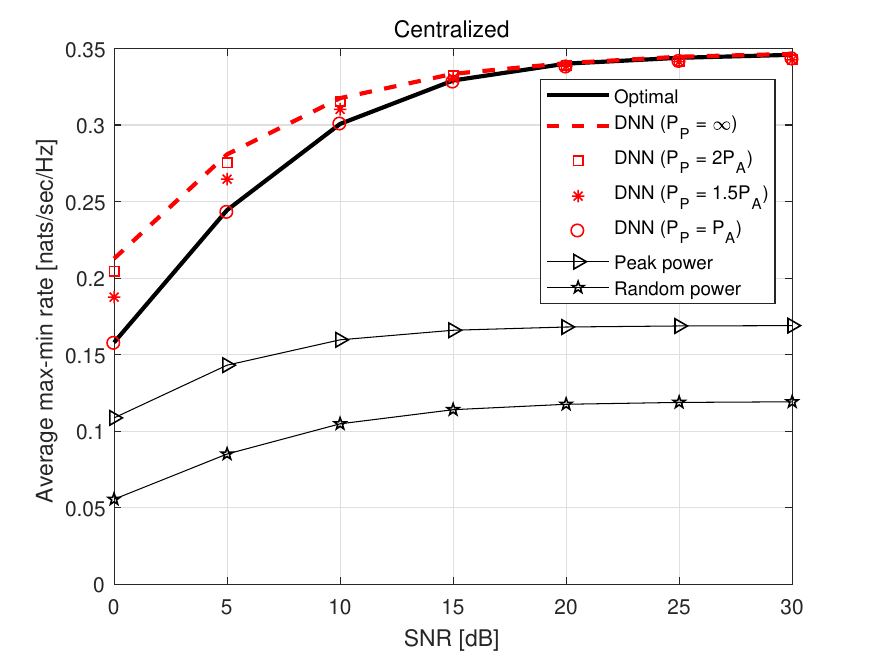}\label{fig:fig9a}
    }
    \subfigure[Average max-min rate with respect to $B$]{
        \includegraphics[width=\linewidth]{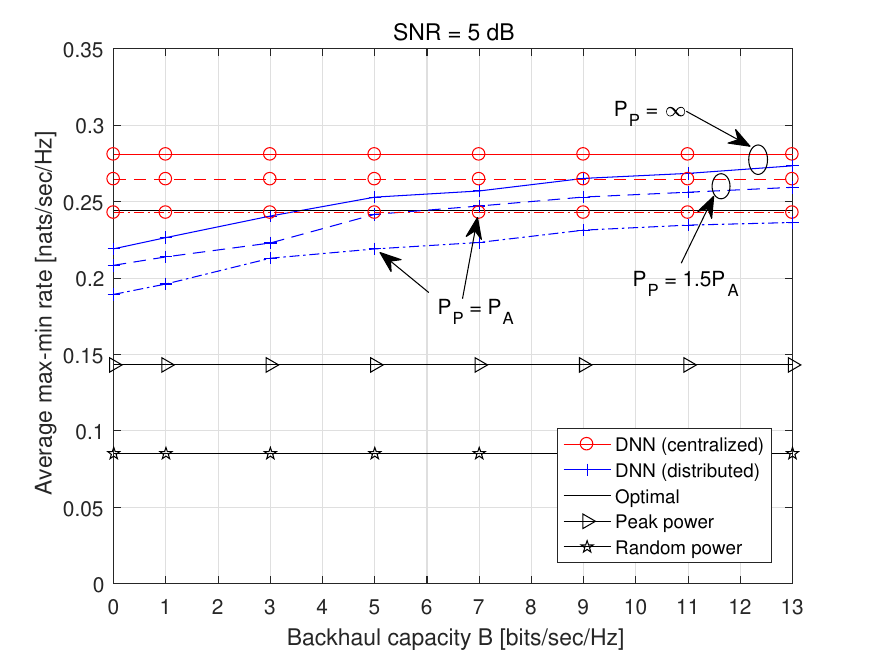}\label{fig:fig9b}
    }
    \caption{Average max-min rate performance of the proposed DL framework with different $P_{P}$.}
    \label{fig:fig9}
\end{figure}

Finally, Fig. \ref{fig:fig9} exhibit the average max-min rate with $P_{A}\leq P_{P}$. The average max-min rate performance of the proposed centralized DL method is shown in Fig. \ref{fig:fig9a} for different $P_{P}$ by changing the SNR. The average max-min rate increases as $P_{P}$ becomes large, and $P_{P}=2P_{A}$ is sufficient to achieve the performance of no bound, i.e., $P_{P}=\infty$. In a high SNR regime, all DNNs with different $P_{P}$ show identical performance to the optimal solution, which is developed for $P_{A}=P_{P}$. This implies that the peak power budget at the transmitter is crucial for the low SNR range. In Fig. \ref{fig:fig9b}, the average max-min rate performance of the proposed distributed approach is illustrated with $\text{SNR}=5\ \text{dB}$ by varying the backhaul link capacity. Similar to the case of $P_{A}=P_{P}$, the performance of the distributed DL method improves as $B$ increases to reach the centralized DNN.

\section{Conclusions and Future Research Directions}\label{sec:sec6}
This work investigates a DL framework that solves a generic constrained optimization associated with wireless networks where multiple nodes are connected capacity-limited backhaul links. To handle inherent non-convexity and distributed nature, this work introduces a DNN to accurately approximate unknown computation rules of the optimal solution. According to the condition of the backhaul capacity, two different scenarios have been considered: For the ideal infinite-capacity case, the centralized DL approach has been provided based on the global information shared among nodes. To consider long-term constraints, a constrained training algorithm has been proposed by applying the Lagrange duality formulation and the primal-dual update rule. In a more realistic finite-capacity backhaul link case, a novel stochastic binarization layer is used to efficiently train the DNNs to avoid a vanishing gradient issue. To validate the efficacy of the proposed DL framework, resource allocation problems in the C-MAC and the IFC are revisited. Intensive numerical results have demonstrated the efficacy of the proposed DL framework for solving an optimization problem in wireless communications.

For future research extensions, several sophisticated applications of the proposed DL framework, such as multi-antenna signal processing applications and wireless resource management in realistic channel models, are worthwhile. The optimization of the DNN structure becomes an essential issue, and the impact of the number and the dimension of hidden layers are to be investigated carefully. Also, feasible applications of the binarization layer can be addressed in mixed integer optimization tasks such as user scheduling and cell association problems. Joint quantization and resource allocation optimization in cloud radio access networks (C-RAN) with the proposed DL framework are worth pursuing. For C-RAN systems, a new DL architecture could be studied with a centralized control unit and distributed mobile devices interconnected via limited-capacity fronthaul and backhaul links.

\nocite{*}
\bibliography{arXiv}
\bibliographystyle{ieeetr}

\end{document}